\documentclass[a4paper,11pt]{article}
\pdfoutput=1
\usepackage{jcappub}
\usepackage[T1]{fontenc}
\usepackage{graphicx}
\DeclareGraphicsExtensions{.pdf}
\usepackage{amsmath,amssymb}
\usepackage{bm}
\usepackage{dcolumn}
\usepackage{xcolor}
\usepackage{placeins}

\newcommand{\dd}{\mathrm{d}}
\newcommand{\rs}{r_{\!*}}
\newcommand{\bc}{b_{c}}
\newcommand{\amax}{\alpha_{\max}}
\newcommand{\rph}{r_{\rm ph}}
\newcommand{\Wodd}{\mathcal{W}^{\rm odd}}
\newcommand{\Weven}{\mathcal{W}^{\rm even}}

\begin{document}

\title{Charge-activated birefringence of an Einstein-tensor-coupled vector field on Reissner--Nordstr\"om black holes}

\author[a]{Mao-Yuan Wan,}
\author[a,1]{Chen Wu}
\affiliation[a]{Xingzhi College, Zhejiang Normal University, Jinhua 321004, Zhejiang, China}
\emailAdd{maoyuan.wan.physics@gmail.com}
\emailAdd{wchenoffd@gmail.com}
\note[1]{Corresponding author.}

\abstract{
We study the propagation of a neutral vector test field non-minimally coupled to the Einstein tensor on a Reissner--Nordstr\"om black hole.
The interaction vanishes on a vacuum background, so electric charge is required to activate the curvature-induced response.
The parity potentials have one symmetry: reversing the sign of the coupling exchanges the electric and magnetic susceptibilities.
Because of this symmetry, the two polarizations respond identically at first order in the coupling. The dominant propagation-level effect is therefore a common shift of the effective photon sphere and of the shadow radius, not a birefringent splitting.
The variation of shadow size can mimic the charge variation of a black hole.
When the coupling is weak, polarization splitting appears only in the second-order term, distinguishing this interaction from corrections involving only charge.
We follow this ordering through full-frequency absorption and scattering, the semiclassical glory, and ray-traced double shadows.
The absorption parity asymmetry and the shadow rim splitting probe the difference between the effective potentials. The polarization-summed backward scattering isolates the helicity-reversing response, which is zero without the coupling.
At the positive-coupling boundary the splitting grows but stays finite. Beyond it the even-parity potential has a pole outside the horizon.
In the high-multipole limit, the two boundaries of the stability window both correspond to the thresholds of bound-state instability, but their physical properties differ: the negatively coupled boundary forms a regular potential well, while the positively coupled boundary marks the breaking of regularity.
A Frobenius analysis shows a forced logarithmic branch at that pole. Without an additional physical connection condition, the continuation across the pole is not unique.
The propagation-level sensitivity is therefore charge-dependent. Separating the dominant common shift from an ordinary change of charge requires polarization information or an independent charge measurement.
}


\maketitle
\flushbottom

\section{Introduction}\label{sec01}

Non-minimal coupling between an electromagnetic field and spacetime curvature arises in low-energy quantum corrections and in many effective field theories. It produces vacuum birefringence: the two polarizations of the photon (or, more generally, of a vector boson) follow different effective metrics~\cite{ref01,ref02}.
The most studied example is the coupling of the field strength to the Weyl tensor $\alpha\,C^{\mu\nu\rho\sigma}F_{\mu\nu}F_{\rho\sigma}$. Since the Weyl tensor equals the Riemann tensor in a vacuum, this coupling is effective on the Schwarzschild background. It decomposes the photon sphere, the shadow, and the wave-optical observables into two branches~\cite{ref03,ref04,ref05,ref06,ref07,ref08,ref09}.
Another, less-studied coupling replaces the Weyl tensor with the Einstein tensor, namely $\alpha\,G^{\mu\rho}F_{\mu\nu}F_{\rho}{}^{\nu}$.
The two couplings differ in one structural respect. In any vacuum solution $G_{\mu\nu}=0$, so the Einstein-tensor term vanishes and the model reduces to Maxwell theory.
An Einstein-tensor-coupled field therefore shows no birefringence on Schwarzschild. A source, in the simplest case a charge, is needed to make $G_{\mu\nu}$ nonzero.
On a Reissner--Nordstr\"om (RN) background, every coupling effect studied below vanishes smoothly as $q\to0$. This limit is a parameter-free consistency test. The Weyl coupling has no such test.

This charge-activated structure was proposed by Chen and Jing~\cite{ref10}, who derived the general wave equation for perturbations of a neutral vector field in RN space, decomposed it into two parity sectors, provided closed-form effective potentials for these two sectors, and investigated the stability of the odd sector in the time domain.
They determined the allowed parameter window from the regularity boundaries of even parity and the thresholds of odd parity dynamical instability.
The only later observational study of this exact model known to us is by Liao, Wei, and Chen~\cite{ref11}. They calculated the gray-body factor of each multipole with a third-order WKB method at the single charge $q=0.2M$.
We extend that work by solving the governing equations numerically over the full frequency range and by scanning the $(\alpha,q)$ plane.
A separate line of work studies scalar fields kinetically coupled to the Einstein tensor. That coupling produces superradiant amplification and gray-body effects~\cite{ref12,ref13}.
Our field is a neutral vector (spin 1, uncharged under the RN-sourcing Maxwell field), so there is no superradiance.
The relevant constraint on the coupling is the stability window discussed below.
Our fixed-background, test-field calculation differs from studies of backreacting non-minimal electromagnetic-curvature couplings and their imaging consequences~\cite{ref14,ref15,ref16,ref17,ref18,ref19}.
We use the effective-field-theory literature on higher-dimension Einstein--Maxwell operators to assess the validity of the coupling, not to select its sign~\cite{ref20,ref21,ref22}.
Tang, Huang, and Zhang recently calculated scalar absorption and scattering on a Frolov black hole~\cite{ref23}. They found that the high-frequency fine structure collapses onto a universal curve set by photon-sphere data, and that it is nearly indistinguishable from a regular metric with the same critical impact parameter.
A vector field has a structure that a scalar field lacks: the photon sphere has two parity branches, and each responds differently to $\alpha$.

Horizon-scale shadows and photon rings are the observables studied in this paper. Shadows were proposed as an observational discriminator before Sgr A* was imaged~\cite{ref24}. Images of Sgr A* and M87*, and polarimetric data on the ring, are now available~\cite{ref25,ref26,ref27}.
This paper reports two results.
The first concerns the potential.
Two independent derivations, using different master variables and gauge choices, give the same closed-form potential.
Its denominator differs by a single sign from Eq.~(16) of Ref.~\cite{ref10}.
Our potential satisfies an exact ``reversal-$\alpha$'' reciprocity relation between the two parities. The form in Eq.~(16) satisfies an ``identical-$\alpha$'' relation instead. This difference changes the dominant phenomenology: the potential derived here produces a first-order parity-\emph{common} shadow shift, while the birefringence splitting appears only at second order.
Reversal-$\alpha$ means that exchanging the electric and magnetic components of the coupled field strength is equivalent to $\alpha\to-\alpha$ (Sec.~\ref{sec02D}). The identical-$\alpha$ relation has no such interpretation.
The second result concerns the ordering of the observables.
Every observable effect still requires both $\alpha\neq0$ and $q\neq0$. The potential derived here separates the observables into a dominant part (the familiar shadow shift, degenerate with an ordinary change of charge) and a subdominant part (the polarization splitting, which is not degenerate with charge).
Breaking the degeneracy therefore requires resolving the second-order splitting or measuring the charge independently.
Vacuum curvature couplings such as the Weyl coupling do not have this second-order, charge-activated behavior, because they remain active on Schwarzschild.

\begin{figure}[t]
  \centering\includegraphics[width=\textwidth]{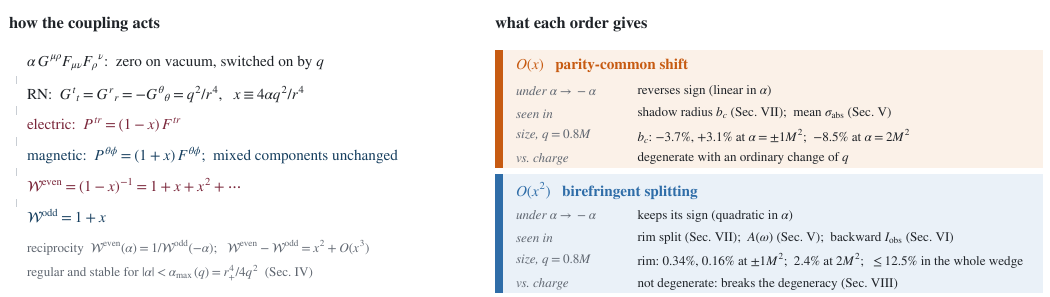}
  \caption{
Schematic of the relation between the coupling, the parity potentials, and the observables of Secs.~\ref{sec05} to~\ref{sec08}.
  The coupling is activated by charge; the two field strength terms determine the parity potential and its reciprocity; at first order in $x=4\alpha q^2/r^4$ the parity potentials shift together, and at second order they split.
  The sizes quoted for each order are the values at the $q=0.8M$ benchmarks of Table~\ref{tab01}; the $12.5\%$ is the largest rim splitting anywhere in the wedge (Sec.~\ref{sec07}).}
  \label{fig01}
\end{figure}

The paper is organized as follows.
Section~\ref{sec02} introduces the action and the master equation, derives the even and odd potentials (for general metric functions) using a gauge-invariant method, and establishes an exact reciprocity relation between the two parities.
Section~\ref{sec03} presents optical metrics and the photon sphere, compares the light cones of the two polarizations, and discusses the validity of this coupling in effective field theory.
Section~\ref{sec04} identifies the stable and regular parameter wedge and examines its positive edge, where a Frobenius analysis of the pole and a bound-state calculation give the growing modes.
Section~\ref{sec05} computes the absorption cross section and its two-order structure.
Section~\ref{sec06} computes the differential scattering cross section, the glory, and the backward birefringence.
Section~\ref{sec07} presents the double-shadow images, rim splitting, and combined benchmark table.
Section~\ref{sec08} discusses the observational reach, the $\alpha$--$q$ degeneracy, and idealized sensitivity curves.
Section~\ref{sec09} concludes.
Figure~\ref{fig01} summarizes the connections between these results.

We use geometrized units $G=c=1$ and the metric signature $(+,-,-,-)$ of Eq.~\eqref{eq03} below, with mass $M$ (set to $1$ unless stated), charge $q$, and coupling $\alpha$.
The signature is that of Ref.~\cite{ref10}, kept so that our action and potentials carry the same signs as in that work.
Our curvature conventions are $R^{\rho}{}_{\sigma\mu\nu}=\partial_\mu\Gamma^{\rho}{}_{\nu\sigma}-\partial_\nu\Gamma^{\rho}{}_{\mu\sigma}+\Gamma^{\rho}{}_{\mu\lambda}\Gamma^{\lambda}{}_{\nu\sigma}-\Gamma^{\rho}{}_{\nu\lambda}\Gamma^{\lambda}{}_{\mu\sigma}$, $R_{\sigma\nu}=R^{\rho}{}_{\sigma\rho\nu}$, and $G_{\mu\nu}=R_{\mu\nu}-\tfrac12 g_{\mu\nu}R$.
Together with the normalizations of $\alpha$ and $q$ in Eqs.~\eqref{eq01} and~\eqref{eq03}, these conventions fix the signs used throughout the text and specify how to translate the expressions to other conventions.

\section{Model and parity potentials}\label{sec02}

\subsection{Action and master equation}\label{sec02A}

The model is a neutral vector field $A_\mu$ with the action
\begin{equation}
  S=\int \dd^4x\sqrt{-g}\left[-\tfrac14\left(F_{\mu\nu}F^{\mu\nu}-4\alpha\,G^{\mu\rho}F_{\mu\nu}F_{\rho}{}^{\nu}\right)\right],
  \label{eq01}
\end{equation}
whose variation gives the modified Maxwell equation
\begin{equation}
  \nabla_\mu\!\left(F^{\mu\nu}-2\alpha\,G^{\mu\rho}F_{\rho}{}^{\nu}+2\alpha\,G^{\nu\rho}F_{\rho}{}^{\mu}\right)=0.
  \label{eq02}
\end{equation}
The field $A_\mu$ is a hidden-sector, dark-photon-like field, distinct from the Maxwell field that sources the RN charge $q$. Identifying the two would modify the background, which would no longer be RN.
The field is treated as a test perturbation on the fixed RN background,
\begin{equation}
  \dd s^2=f\,\dd t^2-f^{-1}\dd r^2-r^2\dd\Omega^2,\qquad f=1-\frac{2M}{r}+\frac{q^2}{r^2},
  \label{eq03}
\end{equation}
with outer horizon $r_+=M+\sqrt{M^2-q^2}$.
On this background $G_{\mu\nu}\neq0$ for $q\neq0$, so the coupling in Eq.~\eqref{eq02} is active. On Schwarzschild ($q=0$), $G_{\mu\nu}\equiv0$ and every term proportional to $\alpha$ vanishes identically. The model therefore reduces exactly to Maxwell theory for any value of $\alpha$.
Unlike the usual $\alpha\to0$ check, the $q\to0$ limit holds for arbitrary $\alpha$, providing a parameter-free consistency check throughout.

Decomposing $A_\mu$ in vector spherical harmonics of odd and even parity and separating variables reduces Eq.~\eqref{eq02} to a single Schr\"odinger-type radial equation per parity ($+$ for even, $-$ for odd) and multipole $\ell$,
\begin{equation}
  \frac{\dd^2\Psi^{\pm}_\ell}{\dd \rs^2}+\left[\omega^2-V^{\pm}(r)\right]\Psi^{\pm}_\ell=0,\qquad \frac{\dd \rs}{\dd r}=\frac{1}{f(r)},
  \label{eq04}
\end{equation}
following the reduction of Ref.~\cite{ref10}, itself built on the odd/even parity split introduced for gravitational perturbations by Regge and Wheeler and by Zerilli~\cite{ref28,ref29} and standard in the black-hole perturbation literature~\cite{ref30,ref31}.
For this model the potentials have the purely multiplicative form
\begin{equation}
  V^{\pm}(r)=f(r)\,\frac{\ell(\ell+1)}{r^2}\,\mathcal{W}^{\pm}(r),
  \label{eq05}
\end{equation}
with no additive term of the kind present in the Weyl-coupled potential~\cite{ref03}. For a general metric function, both parity potentials contain additive contributions (the $h_0,h_1,h_2$ terms of Ref.~\cite{ref10}). These vanish identically for RN and Schwarzschild, so Eq.~\eqref{eq05} is exact for this model (Sec.~\ref{sec02C}).

\subsection{The parity potentials}\label{sec02B}

Reference~\cite{ref10} reports, in our notation,
\begin{equation}
  \Wodd(r)=\frac{r^4+4\alpha q^2}{r^4},
  \label{eq06}
\end{equation}
for the odd parity (their Eq.~(15)), and, for the even parity (their Eq.~(16)),
\begin{equation}
  \Weven_{\rm published}(r)=\frac{r^4}{r^4+4\alpha q^2}=\frac{1}{\Wodd(r)}.
  \label{eq07}
\end{equation}
We re-derived both sectors from the action, Eq.~\eqref{eq01}, using two independent reductions, detailed in the \hyperref[app:parity-derivations]{Appendix}. The first integrates the action over the vector harmonics before varying it, fixes the even-sector angular gauge, and eliminates the remaining potential components through an auxiliary electric momentum. The second varies the four-dimensional action first and reduces its field equations using the gauge-invariant electric momentum $\chi=-r^2P^{tr}$, leaving the gauge freedom unfixed.
The routes share the action, background, and conventions, but use separate symbolic implementations and opposite orders of variation and harmonic reduction. Their canonical variables agree after the constraints have been eliminated.
Both routes reproduce the odd-parity result, Eq.~\eqref{eq06}, exactly.
For the even parity, both routes give
\begin{equation}
  V_{\rm even}(r)=f(r)\,\frac{\ell(\ell+1)\,r^2}{r^4-4\alpha q^2},
  \label{eq08}
\end{equation}
i.e., $\Weven(r)=r^4/(r^4-4\alpha q^2)$, which differs from Eq.~\eqref{eq07} by a single sign in the denominator. 
Equivalently, $\Weven_{\rm published}(\alpha)=\Weven(-\alpha)$. 
The difference is the sign of the coupling relative to the odd sector. The component structure of Eq.~\eqref{eq10} fixes this sign before any master variable is chosen. Both forms use the tortoise coordinate of Eq.~\eqref{eq04} and the Schr\"odinger normalization with unit coefficient of $\omega^2$, so they can be compared directly.
Two facts fix Eq.~\eqref{eq08} unambiguously.
First, for a given master variable, here the gauge-invariant momentum $\chi$ of Sec.~\ref{sec02C1}, the potential is read from the Schr\"odinger form in the tortoise coordinate, which has no first-derivative term. Any $r$-dependent rescaling of $\chi$ reintroduces a first-derivative term. The potential of $\chi$ is therefore fixed up to a constant rescaling.
Second, the coupled field strength, Eq.~\eqref{eq10}, fixes the signs before any master variable is chosen: the electric component is rescaled by $1-4\alpha q^2/r^4$, the magnetic component by $1+4\alpha q^2/r^4$, and the even potential carries the inverse of its component factor.
The two routes relate to Eq.~\eqref{eq08} as follows.
The gauge-invariant route of Sec.~\ref{sec02C} gives Eq.~\eqref{eq08} symbolically.
In the first route, the auxiliary momentum becomes $p=r^2\mathcal A F_{tr}$, where $F_{tr}$ here denotes its radial harmonic coefficient and $\mathcal A=1-4\alpha q^2/r^4$. The same result follows from the intermediate $A_r=hY e^{-i\omega t}$ equation: its unnormalized form contains both a first derivative and the geometric terms $ff''+(f')^2$. For $\omega\ne0$, the transformation $u=fh=i\omega p/\lambda$ cancels these terms exactly and yields Eq.~\eqref{eq08}; the momentum formulation also covers $\omega=0$. Section~\ref{app:rn-specialization} of the Appendix gives the steps. The harmonic ansatz is the same as in Ref.~\cite{ref10}, but the literal printed even master variable is not our canonical variable, as shown in Sec.~\ref{app:published-variable} of the Appendix.
All numerical results in this paper are computed with Eq.~\eqref{eq08}.
Equation~\eqref{eq06} is unchanged: our methods differ from Ref.~\cite{ref10} only in the even-parity potential, Eq.~\eqref{eq08}.

The sign of the coupling term in Eq.~\eqref{eq08} follows from the RN geometry.
The Einstein tensor satisfies $G^{t}{}_{t}=G^{r}{}_{r}=-G^{\theta}{}_{\theta}$ on RN. The two Einstein-tensor insertions in the electric, $(t,r)$ component momentum therefore add with the same sign. They give the factor $\left(1-4\alpha q^2/r^4\right)$ in $P^{tr}=-F_{tr}\left(1-4\alpha q^2/r^4\right)$, reducing $P^{tr}$ for $\alpha>0$ rather than enhancing it.
The same pair of insertions cancels identically in every mixed time--angular and radial--angular component, and combines with the opposite sign in the purely angular $(\theta,\phi)$ component, producing the factor $\left(1+4\alpha q^2/r^4\right)$ of Eq.~\eqref{eq06} in the odd-parity (magnetic-type) sector.
The $W$-factor structure of Eq.~\eqref{eq05} is therefore special to RN (and Schwarzschild). For a general metric function $f$ containing, for example, a $c_3/r^3$ term, the non-$\ell(\ell+1)$ part of the even-parity potential is nonzero. The reciprocity relation derived below thus does not hold for a generic background or coupling.
Section~\ref{sec02C} gives the step-by-step derivation of both parity potentials, Eqs.~\eqref{eq06} and~\eqref{eq08}, along the gauge-invariant route, from the modified Maxwell equation, Eq.~\eqref{eq02}, to the potentials.
Equation~\eqref{eq13} fixes the sign and can be checked directly, independently of the numerical comparison. Section~\ref{sec02C2} gives the general-$f$ potentials.
The same derivation gives the general-$f$ structure written in Sec.~\ref{sec02C2}. Each auxiliary term follows from varying Eq.~\eqref{eq01} and requiring every coupling-dependent term to vanish at $q=0$.

\subsection{Detailed derivation of the master equations and the general-$f$ potentials}\label{sec02C}

\subsubsection{Gauge-invariant derivation of the parity master equations}\label{sec02C1}

Here we give the second derivation described in Sec.~\ref{sec02B}. Starting from the modified Maxwell equation, Eq.~\eqref{eq02}, we obtain the master equations of both parity sectors, Eqs.~\eqref{eq06} and~\eqref{eq08}.
The derivation identifies the step that fixes the sign of the $4\alpha q^2/r^4$ term.

On the RN background, Eq.~\eqref{eq03}, the mixed Einstein tensor is diagonal, with
\begin{equation}
  G^{t}{}_{t}=G^{r}{}_{r}=\frac{q^2}{r^4},\qquad G^{\theta}{}_{\theta}=G^{\phi}{}_{\phi}=-\frac{q^2}{r^4},
  \label{eq09}
\end{equation}
as verified by direct calculation of the Christoffel symbols, Ricci tensor, and Einstein tensor. The $(t,r)$ and angular components have curvature of equal magnitude and opposite sign.

Since the mixed Einstein tensor is diagonal, the modified field strength tensor $P^{\mu\nu}=F^{\mu\nu}-2\alpha G^{\mu\rho}F_{\rho}{}^{\nu}+2\alpha G^{\nu\rho}F_{\rho}{}^{\mu}$ in Eq.~\eqref{eq02} has a closed form. The two contractions give $G^{\mu\rho}F_{\rho}{}^{\nu}=G^{\mu}{}_{\mu}F^{\mu\nu}$ and $G^{\nu\rho}F_{\rho}{}^{\mu}=-G^{\nu}{}_{\nu}F^{\mu\nu}$ (no sums), hence
\begin{equation}
  P^{\mu\nu}=\left[1-2\alpha\left(G^{\mu}{}_{\mu}+G^{\nu}{}_{\nu}\right)\right]F^{\mu\nu}\qquad\text{(no sum)}.
  \label{eq10}
\end{equation}
Together with Eq.~\eqref{eq09}, this determines all components. The electric $(t,r)$ component has $G^{t}{}_{t}+G^{r}{}_{r}=2q^2/r^4$ and is rescaled by $1-4\alpha q^2/r^4$. The purely angular, magnetic $(\theta,\phi)$ component has $G^{\theta}{}_{\theta}+G^{\phi}{}_{\phi}=-2q^2/r^4$ and is rescaled by $1+4\alpha q^2/r^4$. The mixed time--angular and radial--angular components, $(t,\theta)$, $(t,\phi)$, $(r,\theta)$ and $(r,\phi)$, satisfy $G^{\mu}{}_{\mu}+G^{\nu}{}_{\nu}=0$ and remain unchanged.
The even (electric-type) sector depends on the $(t,r)$ component, and the odd (magnetic-type) sector on the $(\theta,\phi)$ component. The factors $1\mp4\alpha q^2/r^4$, and hence the exact reciprocity of Sec.~\ref{sec02D}, are fixed before choosing a master variable.

For the even-parity, axisymmetric ($m=0$) sector at fixed $\ell$, write
\begin{equation}
  \begin{aligned}
    A_t&=j(r)Y e^{-i\omega t},\qquad A_r=h(r)Y e^{-i\omega t},\\
    A_\theta&=k(r)\,\partial_\theta Y\, e^{-i\omega t},
  \end{aligned}
  \label{eq11}
\end{equation}
with $Y=P_\ell(\cos\theta)$ and $A_\phi=0$, leaving the gauge freedom $k(r)$ unfixed rather than setting $k=0$ as in the first route in Sec.~\ref{app:action-route} of the Appendix.
The gauge-invariant combinations
\begin{equation}
  \Phi\equiv j+i\omega k,\qquad Z\equiv h-k',\qquad E_r\equiv -i\omega h-j'
  \label{eq12}
\end{equation}
(with $E_r=F^{\rm rad}_{tr}$, the radial electric-field component) satisfy the Bianchi identity $E_r=-(\Phi'+i\omega Z)$. Direct substitution shows that every field equation below is independent of $k$ when expressed in $\Phi,Z,E_r$, verifying gauge invariance explicitly.

The step that fixes the sign in this sector is the electric, $(t,r)$ component of Eq.~\eqref{eq10},
\begin{equation}
  P^{tr}=-F_{tr}\left(1-\frac{4\alpha q^2}{r^4}\right),
  \label{eq13}
\end{equation}
in which the coupling contribution has the same sign as $F_{tr}$ (reducing $|P^{tr}|$ for $\alpha>0$). Since $G^{t}{}_{t}=G^{r}{}_{r}$, Eq.~\eqref{eq09}, the two insertions add rather than cancel.
Of the field strengths built from $(A_t,A_r,A_\theta)$, only $F_{tr}$ is modified: the $(t,\theta)$ and $(r,\theta)$ components entering this sector have $G^{\mu}{}_{\mu}+G^{\nu}{}_{\nu}=0$ in Eq.~\eqref{eq10} and carry no $\alpha$-dependence at all.
This structure fixes the sign of the coupling term in the even-parity potential.

Defining the gauge-invariant electric momentum
\begin{equation}
  \chi\equiv -r^2P^{tr}=r^2\left(1-\frac{4\alpha q^2}{r^4}\right)E_r,
  \label{eq14}
\end{equation}
the three radial field equations ($\nu=t,r,\theta$ components of $\nabla_\mu P^{\mu\nu}=(1/\sqrt{-g})\partial_\mu(\sqrt{-g}P^{\mu\nu})=0$), after dividing out the common angular factor and an overall $1/r^2$, reduce to
\begin{align}
  \nu=r:&\quad i\omega\chi-\lambda f Z=0,\label{eq15}\\
  \nu=\theta:&\quad (fZ)'=-\frac{i\omega\Phi}{f},\label{eq16}\\
  \nu=t:&\quad \chi'+\frac{\lambda\Phi}{f}=0,
  \label{eq17}
\end{align}
with $\lambda\equiv\ell(\ell+1)$. Once $Z$ and $\Phi$ are eliminated through the $\nu=r$ and $\nu=t$ equations, the $\nu=\theta$ equation is satisfied identically at every $\omega$, including $\omega=0$; for $\omega\neq0$ the $\nu=t$ (Gauss) equation is in turn implied by the other two and the Bianchi relation, as verified symbolically for $\ell=1,2,3$.
We eliminate $Z=i\omega\chi/(\lambda f)$ and $\Phi=-(f/\lambda)\chi'$ from the $\nu=r$ and $\nu=t$ equations and substitute them into the Bianchi relation $E_r=-(\Phi'+i\omega Z)$. Using $E_r=\chi/[r^2(1-4\alpha q^2/r^4)]$ from the definition of $\chi$ gives a second-order equation for $\chi(r)$. Conversion to the tortoise coordinate removes the first-derivative term, as verified symbolically. Reading the potential from $\dd^2\chi/\dd \rs^2+[\omega^2-V]\chi=0$ then gives
\begin{equation}
  V_{\rm even}(r)=\frac{\lambda f}{r^2\left(1-4\alpha q^2/r^4\right)}=\frac{f\,\lambda\,r^2}{r^4-4\alpha q^2},
  \label{eq18}
\end{equation}
identical to Eq.~\eqref{eq08}, with a denominator that is manifestly $\omega$-independent and exactly linear in $\alpha q^2$.
The $\alpha=0$ limit of Eq.~\eqref{eq18} is exactly $f\lambda/r^2$, the Maxwell-on-RN potential.

The odd-parity sector follows from the same structure.
With the odd-parity ansatz and $A_t=A_r=A_\theta=0$, the only nonvanishing field strengths are
\begin{align}
 F_{t\phi}&=-i\omega a\sin\theta\,\partial_\theta Y,\nonumber\\
 F_{r\phi}&=a'\sin\theta\,\partial_\theta Y,\nonumber\\
 F_{\theta\phi}&=-\lambda a\sin\theta\,Y.
\end{align}
By Eq.~\eqref{eq10}, the first two are unchanged and the purely angular magnetic component is rescaled, $P^{\theta\phi}=\left(1+4\alpha q^2/r^4\right)F^{\theta\phi}$.
In the single dynamical equation, the $\nu=\phi$ component of $\nabla_\mu P^{\mu\phi}=0$, the $\omega^2$ term and the radial derivative term are therefore the same as in the Maxwell case. The coupling enters only the centrifugal term, through the rescaled $(\theta,\phi)$ component:
\begin{equation}
  V_{\rm odd}(r)=\frac{f\,\lambda}{r^2}\left(1+\frac{4\alpha q^2}{r^4}\right),
  \label{eq19}
\end{equation}
which is Eq.~\eqref{eq06}, verified symbolically for $\ell=1,2$. Agreement between the two routes in this sector checks the sign and normalization conventions used above.
The factors $1\mp4\alpha q^2/r^4$ thus follow from applying the same calculation to the two nontrivial components of Eq.~\eqref{eq10}.

\subsubsection{General-$f$ potentials from an independent derivation}\label{sec02C2}

For a general static, spherically symmetric metric function $f(r)$, the reduction of Sec.~\ref{sec02C1} goes through with the factors of Eq.~\eqref{eq10} replaced by $\mathcal{A}=1+4\alpha(rf'+f-1)/r^2$ for the electric component, $\mathcal{B}=1+2\alpha f''+4\alpha f'/r$ for the magnetic component, and $\mathcal{D}=1+\alpha(r^2f''+4rf'+2f-2)/r^2$ for the mixed $(t,a)$ and $(r,a)$, $a=\theta,\phi$, which no longer equals unity.
The radial equations then carry first-derivative terms proportional to $\dot{\mathcal D}/\mathcal D$, and the Schr\"odinger normal forms for $\Psi^{\rm odd}=\sqrt{\mathcal D}\,a$ and $\Psi^{\rm even}=\chi/\sqrt{\mathcal D}$ read
\begin{equation}
  V^{\rm odd}=\frac{f\lambda}{r^2}\frac{\mathcal B}{\mathcal D}+\frac{\ddot{\sqrt{\mathcal D}}}{\sqrt{\mathcal D}},\qquad
  V^{\rm even}=\frac{f\lambda}{r^2}\frac{\mathcal D}{\mathcal A}+\sqrt{\mathcal D}\;\ddot{\bigl(\mathcal D^{-1/2}\bigr)},
  \label{eq20}
\end{equation}
with dots denoting $\dd/\dd\rs$, valid wherever $\mathcal A,\mathcal D\neq0$, and verified symbolically.
For RN the identity
\begin{equation}
  r^2f''+4rf'+2f-2\equiv0,
  \label{eq21}
\end{equation}
satisfied by $f=1-2M/r+q^2/r^2$ for any $M,q$, gives $\mathcal D\equiv1$: the additive terms vanish and the multiplicative factors reduce to Eqs.~\eqref{eq06} and~\eqref{eq08}.
The general-$f$ potentials quoted in Ref.~\cite{ref10} (their Eqs.~(9)--(10)) reduce through the same identity to Eq.~\eqref{eq06} and to Eq.~\eqref{eq07}: their general-$f$ even denominator is $1-4\alpha(rf'+f-1)/r^2$ where ours is $\mathcal A$, so the sign difference between Eqs.~\eqref{eq07} and~\eqref{eq08} is not introduced by the specialization to RN.
For a general $f(r)$ with additional terms, such as $c_3/r^3$, $\mathcal D\neq1$ and the additive terms are nonzero. The multiplicative structure of Eq.~\eqref{eq05}, and with it the exact reciprocity of Eq.~\eqref{eq22}, is then special to RN (and Schwarzschild). The reciprocity holds for this model and background, not in general.

\subsection{Exact reciprocity}\label{sec02D}

\begin{table}
\caption{Photon-sphere and shadow benchmarks ($M=1$; $q$ in units of $M$, $\alpha$ in units of $M^2$). ``odd'' and ``even'' refer to the two optical metrics; $\alpha_{\max}(q)=r_+^4/(4q^2)$ is the wedge half-width of Sec.~\ref{sec04A}; $\eta_c=\Lambda_c/\Omega_c$ is the ratio of the photon orbit's Lyapunov exponent to its orbital frequency (Sec.~\ref{sec05}). Rim splitting is $(b_c^{\rm odd}-b_c^{\rm even})/b_c(\alpha{=}0)$ and shadow shift is $[\tfrac12(b_c^{\rm odd}+b_c^{\rm even})-b_c(\alpha{=}0)]/b_c(\alpha{=}0)$, in percent, relative to the same-$q$, $\alpha{=}0$ reference. Columns 3--7 carry four decimals except in the $q\to0$ rows, enough to resolve the sub-percent splittings; the $\alpha{=}0$ rows are exact-RN controls and the $q\to0$ rows the Schwarzschild limit.}
\label{tab01}
\renewcommand{\arraystretch}{0.85}

\resizebox{\textwidth}{!}{%
\begin{tabular}{ccccccccccc}
$q$ & $\alpha$ & $\alpha_{\max}$ & $r_{ph}^{\rm odd}$ & $r_{ph}^{\rm even}$ & $b_c^{\rm odd}$ & $b_c^{\rm even}$ & $\eta_c^{\rm odd}$ & $\eta_c^{\rm even}$ & split (\%) & shift (\%) \\
\hline
\multicolumn{11}{l}{\textit{In-wedge benchmarks}} \\
$0.8$ & $1$ & $2.5600$ & $2.3618$ & $2.3378$ & $4.3860$ & $4.3705$ & $0.835$ & $0.810$ & $0.341$ & $-3.69$ \\
$0.8$ & $-1$ & $2.5600$ & $2.6094$ & $2.5970$ & $4.6908$ & $4.6835$ & $0.959$ & $0.945$ & $0.160$ & $3.11$ \\
$0.6$ & $2$ & $7.2900$ & $2.6404$ & $2.6277$ & $4.7301$ & $4.7216$ & $0.911$ & $0.898$ & $0.174$ & $-2.73$ \\
$0.8$ & $2$ & $2.5600$ & $2.2540$ & $2.1021$ & $4.2147$ & $4.1057$ & $0.804$ & $0.650$ & $2.399$ & $-8.49$ \\
\multicolumn{11}{l}{\textit{$\alpha=0$ references (exact-RN control)}} \\
$0.2$ & $0$ & $96.0200$ & $2.9731$ & $2.9731$ & $5.1612$ & $5.1612$ & $0.995$ & $0.995$ & $0$ & $0$ \\
$0.4$ & $0$ & $21.0799$ & $2.8892$ & $2.8892$ & $5.0530$ & $5.0530$ & $0.981$ & $0.981$ & $0$ & $0$ \\
$0.6$ & $0$ & $7.2900$ & $2.7369$ & $2.7369$ & $4.8587$ & $4.8587$ & $0.951$ & $0.951$ & $0$ & $0$ \\
$0.8$ & $0$ & $2.5600$ & $2.4849$ & $2.4849$ & $4.5460$ & $4.5460$ & $0.890$ & $0.890$ & $0$ & $0$ \\
$0.95$ & $0$ & $0.8214$ & $2.1671$ & $2.1671$ & $4.1762$ & $4.1762$ & $0.785$ & $0.785$ & $0$ & $0$ \\
\multicolumn{11}{l}{\textit{$q\to0$ Schwarzschild anchor}} \\
$\to0$ & $2$ & -- & $3$ & $3$ & $5.196$ & $5.196$ & $1.000$ & $1.000$ & $0$ & $\sim-0$ \\
$\to0$ & $-2$ & -- & $3$ & $3$ & $5.196$ & $5.196$ & $1.000$ & $1.000$ & $0$ & $\sim+0$ \\
\end{tabular}%
}

\vspace{-8pt}
\end{table}

\begin{figure}[t]
  \centering
  \includegraphics[width=\textwidth]{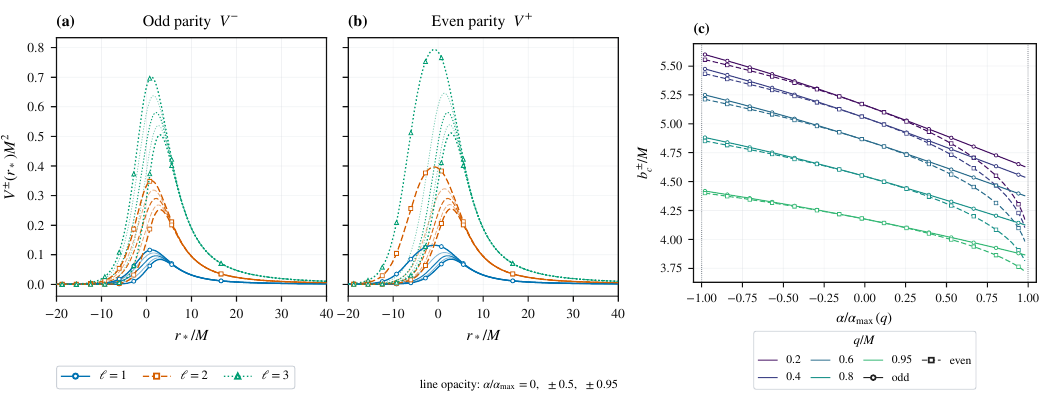}
  \caption{
  Parity potentials and critical impact parameters over the wedge, normalized by the half-width $\amax(q)=r_+^4/4q^2$ of Sec.~\ref{sec04A}.
  (a),(b) Odd- and even-parity potentials $V^{\rm odd}(r_*)$ and $V^{\rm even}(r_*)$ at $q=0.8M$ for $\ell=1,2,3$ and $\alpha/\amax=0,\,\pm0.5,\,\pm0.95$; color, line style, and markers distinguish $\ell$, and opacity and line weight the coupling magnitude.
  (c) $\bc^{\pm}$ versus $\alpha/\amax(q)$ for $q/M=0.2,0.4,0.6,0.8,0.95$; color distinguishes charge and line style parity.}
  \label{fig02}
  \vspace{-8pt}
\end{figure}

The potentials derived here satisfy an exact reciprocity relation that Eq.~\eqref{eq07} does not:
\begin{equation}
  \Weven(\alpha)=\frac{1}{\Wodd(-\alpha)}\quad\text{at every }r,
  \label{eq22}
\end{equation}
equivalently $H^{\rm odd}(-\alpha)\,H^{\rm even}(\alpha)=(f/r^2)^2$ for the optical potentials $H^{\pm}=f\,\mathcal{W}^{\pm}/r^2$ of Sec.~\ref{sec03A}.
Equation~\eqref{eq22} follows algebraically from Sec.~\ref{sec02B}.
The coupling multiplies the electric, $(t,r)$ momentum by $(1-4\alpha q^2/r^4)$ and the magnetic, $(\theta,\phi)$ by $(1+4\alpha q^2/r^4)$. Exchanging these components is equivalent to reversing the sign of $\alpha$. The even potential contains the inverse of its factor, whereas the odd potential contains the factor itself, giving Eq.~\eqref{eq22}.
Under $F\to{}^\star F$ the Maxwell term of Eq.~\eqref{eq01} reverses sign, since $({}^\star F)_{\alpha\beta}({}^\star F)^{\alpha\beta}=-F_{\alpha\beta}F^{\alpha\beta}$, whereas the coupling term is unchanged, by the identity $({}^\star F)_{\mu\alpha}({}^\star F)_{\nu}{}^{\alpha}=F_{\mu\alpha}F_{\nu}{}^{\alpha}-\tfrac12 g_{\mu\nu}F_{\alpha\beta}F^{\alpha\beta}$ together with the tracelessness $G^{\mu}{}_{\mu}=0$ of the Einstein tensor on RN. Writing $\mathcal{L}_\alpha[F]$ for the integrand of Eq.~\eqref{eq01}, $\mathcal{L}_\alpha[{}^\star F]=-\mathcal{L}_{-\alpha}[F]$. Relative to the Maxwell term, the coupling reverses sign. The electric$\leftrightarrow$magnetic exchange is therefore $\alpha\to-\alpha$ on any Ricci-traceless background, and on no other. This is a second reason why the reciprocity is special to RN.
This identity must be distinguished from a duality symmetry.
The coupled theory is not invariant under electric--magnetic duality (the product of the two factors is $1-16\alpha^2q^4/r^8$, not unity), so Eq.~\eqref{eq22} is not a symmetry of the equations of motion but an exact identity between the two sectors of the same theory.
The same-$\alpha$ relation $\Weven=1/\Wodd(+\alpha)$ satisfied by Eq.~\eqref{eq07} is also an identity between two functions; it is not, however, the one in which the factor exchange appears as a sign reversal of the coupling.
Writing $x\equiv4\alpha q^2/r^4$, the two eikonal factors are
\begin{equation}
  \begin{aligned}
    \Wodd&=1+x,\\
    \Weven&=\frac{1}{1-x}=1+x+x^2+\cdots,
  \end{aligned}
  \label{eq23}
\end{equation}
the second an exact geometric series inside its radius of convergence, $|x|<1$.
This resummed form explains the change in phenomenology described in the introduction. At first order in $x$, $\Wodd$ and $\Weven$ coincide, so the leading effect is a parity-\emph{common} shift of the effective potential. The splitting, $\Weven-\Wodd=x^2+O(x^3)$, enters only at second order.
For the alternative same-$\alpha$ form, $\Weven_{\rm published}=1/(1+x)=1-x+x^2-\cdots$, the parities would differ at first order, with $\Wodd$ and this form shifting in opposite directions. The potential derived here does not produce that first-order birefringence.

\section{Optical metrics, causality and effective-field-theory validity}\label{sec03}

\subsection{Optical metrics and photon spheres}\label{sec03A}

Photons of each parity follow null geodesics of a parity-dependent optical metric,
\pagebreak[3]
\begin{equation}
  \dd s_{\pm}^2=f\,\dd t^2-f^{-1}\dd r^2-\frac{r^2}{\mathcal{W}^{\pm}(r)}\,\dd\Omega^2,
  \label{eq24}
\end{equation}
whose angular part carries the factor $1/\mathcal{W}^{\pm}$. This reproduces the geometric-optics (eikonal) limit of Eq.~\eqref{eq04}.
The associated radial potential is $H^{\pm}(r)=f(r)\,\mathcal{W}^{\pm}(r)/r^2$, with photon sphere $\rph^\pm$ at $\dd H^\pm/\dd r=0$ and critical impact parameter
\begin{equation}
  \bc^{\pm}=\left[H^{\pm}(\rph^\pm)\right]^{-1/2}.
  \label{eq25}
\end{equation}
Ray tracing and the geodesic scan use these optical metrics rather than the RN metric. The photon sphere and the shadow edge are therefore computed from the same potential as the wave equation, Eq.~\eqref{eq05}.
As $q\to0$, both critical impact parameters approach the Schwarzschild value $\bc=3\sqrt3\,M$ for any $\alpha$ (Table~\ref{tab01}), directly testing the $q$-activation structure. The same limit holds at the wave level: the odd and even absorption cross sections in Sec.~\ref{sec05}, computed at $q=10^{-6}M$ and $\alpha=\pm2M^2$, coincide to machine precision over $1\le M\omega\le25$.
Figure~\ref{fig02} summarizes both quantities over the parameter wedge of Sec.~\ref{sec04A}: panels (a) and (b) show the odd and even potentials at fixed charge, and panel (c) shows the critical impact parameter of Eq.~\eqref{eq25} against the normalized coupling for five charges.

\subsection{Causality and light cones}\label{sec03B}

\begin{figure}[t]
  \centering\includegraphics[width=\columnwidth]{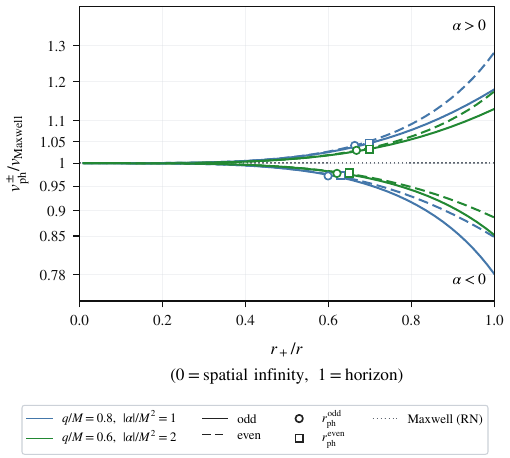}
  \caption{
  Local angular phase velocity of each parity relative to the Maxwell (RN) light cone, $v_{\rm ph}^{\pm}/v_{\rm Maxwell}=\sqrt{\mathcal{W}^{\pm}(r)}$, on a logarithmic ordinate, for $q/M=0.8$, $|\alpha|/M^2=1$ ($|\alpha|/\amax=0.39$) and $q/M=0.6$, $|\alpha|/M^2=2$ ($|\alpha|/\amax=0.27$) at both signs of the coupling.
  The $x$-axis is normalized to $r_+/r$, with the event horizon at $r_+/r=1$ and the spatial infinity at $r_+/r=0$. Open markers mark the photon sphere of each parity.}
  \label{fig03}
\end{figure}

Because $\mathcal{W}^{\rm odd}\neq\mathcal{W}^{\rm even}$ for $\alpha\neq0$, the polarizations propagate on distinct light cones. We examine whether either permits propagation faster than the metric (RN) light cone in the sense of Drummond and Hathrell~\cite{ref01}.
Figure~\ref{fig03} shows the local angular phase velocity of each parity relative to the RN (Maxwell) cone, $v_{\rm ph}^{\pm}(r)/v_{\rm Maxwell}(r)=\sqrt{\mathcal{W}^\pm(r)}$, for representative points inside the wedge and both signs of $\alpha$.
Both parities shift to the \emph{same} side of the Maxwell cone at first order in $\alpha$: for $\alpha>0$ both ratios exceed unity near the horizon (up to about $25\%$ at the benchmark points shown), and for $\alpha<0$ both fall below it.
The two parities separate from each other only at second order.
The two signs of the coupling are related by the reciprocity of Eq.~\eqref{eq22}, $v_{\rm ph}^{\rm odd}(-\alpha)\,v_{\rm ph}^{\rm even}(+\alpha)=1$, so on the logarithmic ordinate of Fig.~\ref{fig03} the $\alpha<0$ curves are the mirror images of the $\alpha>0$ curves through the Maxwell line.
This is the same qualitative phenomenon as the Drummond--Hathrell effect on Schwarzschild~\cite{ref01}.
A curvature-induced correction can make one or both polarizations locally superluminal relative to the background metric without producing closed causal curves.
Inside the wedge, both $\mathcal{W}^{\pm}$ are positive everywhere outside the horizon. Each optical metric in Eq.~\eqref{eq24} is therefore static and Lorentzian, with the same Killing time $t$. This coordinate is a common time function for both effective cones, so each effective geometry is stably causal.
Causality is properly assessed with respect to the wider of the two effective cones rather than the metric cone itself~\cite{ref32,ref33}.
The local superluminality is causally benign in the sense discussed in a recent reassessment of the Drummond--Hathrell effect~\cite{ref34}.

\subsection{Effective-field-theory validity and the sign of $\alpha$}\label{sec03C}

We now ask whether positivity or causality bounds fix the sign or magnitude of $\alpha$.
The Einstein tensor $G_{\mu\nu}=R_{\mu\nu}-\tfrac12 g_{\mu\nu}R$ is built from the Ricci tensor alone and has no Weyl part. On shell (using $G_{\mu\nu}=8\pi T_{\mu\nu}$) a field redefinition therefore turns our operator into a $T\cdot F\cdot F$ contact term of portal type. For this reason, published dimension-eight Einstein--Maxwell operator bases keep only $(F^2)^2$, $\mathrm{Tr}\,F^4$, and Riemann--$FF$ terms, and no independent Ricci--$FF$ operator~\cite{ref20,ref21}.
This on-shell reduction concerns a backreacting theory and does not directly apply to our fixed-background, test-field calculation, which neglects backreaction. Here the coupling is a phenomenological probe.
Multi-field positivity bounds constrain portal-type coefficients only in magnitude, through a Cauchy--Schwarz-type argument against an unknown diagonal term, not in sign~\cite{ref35,ref36}. Causality bounds of the Camanho--Edelstein--Maldacena--Zhiboedov (CEMZ) type change sign with the charge and polarization of the probe~\cite{ref37}. The Drummond--Hathrell-type superluminality discussed above is sign-independent and causally benign rather than excluded~\cite{ref34}.
None of these arguments selects a sign of $\alpha$. We therefore constrain the coupling through stability, regularity, and observational assumptions.
We present results mainly within the range $|\alpha|/r_+^2 \lesssim 1$.
This ceiling is a dimensional estimate for a length-squared coupling at a horizon of curvature radius $r_+$, not a bound derived from a specific ultraviolet completion. The local expansion parameter $|\alpha|q^2/r^4$ is smaller everywhere outside the horizon.
The mathematically regular window of Sec.~\ref{sec04A} is where the potentials have no exterior pole, and the dynamically stable part of it is where no bound-state instability exists.
Within the controlled region $|\alpha|\lesssim r_+^2$, the coupling term is the leading term of a controlled expansion.

\section{The parameter window and its edges}\label{sec04}

\subsection{The stable and regular wedge}\label{sec04A}

\begin{figure}[t]
  \centering\includegraphics[width=\textwidth]{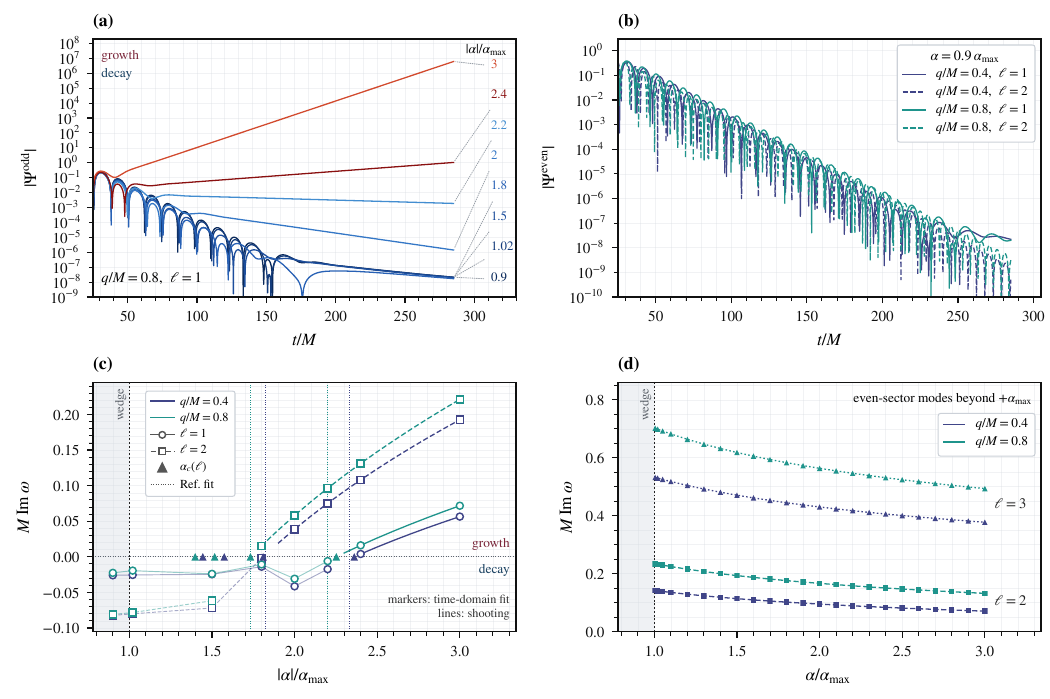}
  \caption{
  (a) Negative (odd-parity) edge: $|\Psi^{\rm odd}(t)|$ from time-domain evolution at several $|\alpha|/\amax$.
  (b) Positive (even-parity) edge: $|\Psi^{\rm even}(t)|$ at $\alpha=0.9\,\amax$ for four $(q,\ell)$ combinations.
  (c) Growth rates $M\,\mathrm{Im}\,\omega$ at the negative edge for $q/M=0.4,0.8$ and $\ell=1,2$: markers are the slopes fitted to evolutions like those in (a), lines the odd-sector bound-state rates from the shooting method, and faint segments join the decaying points below threshold; triangles on the zero line are the thresholds $\alpha_c(\ell)$ for $\ell=1,\dots,4$ (right to left), and dotted verticals the thresholds of Ref.~\cite{ref10} in our units.
  (d) Growth rates of the even-sector bound states beyond the positive edge, $\ell=2,3$ at $q/M=0.4,0.8$, for $\alpha$ from $1.005$ to $3\,\amax$.
  Both edges of the stable and regular wedge, $q=0.8$, $\ell=1$ unless noted; the wedge $|\alpha|<\amax$ is shaded in (c) and (d).}
  \label{fig04}
\end{figure}

The potentials bound the coupling from both sides, and the two edges differ in physical character.
For sufficiently negative $\alpha$ the odd-parity potential turns negative near the horizon and binds a state with $\omega^2<0$ beyond a finite-$\ell$ threshold $\alpha_c(\ell)<0$. Because the odd potential is proportional to $\ell(\ell+1)$, the coupling at which it first turns negative is the same for every multipole, and the finite-$\ell$ thresholds descend toward that value as $\ell$ grows.
Computed directly with the shooting method of Sec.~\ref{sec04B}, applied here to a regular well, $\alpha_c(\ell{=}1,\dots,4)=-49.8$, $-38.1$, $-33.2$, $-30.5$ at $q=0.4M$ and $-5.77$, $-4.44$, $-3.88$, $-3.58$ at $q=0.8M$; each lies beyond the envelope $\amax(q)$ and the ratio $|\alpha_c(\ell)|/\amax$ decreases toward unity with $\ell$, as the eikonal argument requires.
The fit of Ref.~\cite{ref10} (their Eqs.~(21)--(22)), converted to our units, gives $\alpha_c(\ell{=}1)\simeq-49.2$ and $\alpha_c(\ell{=}2)\simeq-38.4$ at $q=0.4$, and $-5.63$ and $-4.44$ at $q=0.8$, within $2.4\%$ of ours; extrapolated to $\ell\to\infty$ it gives $-21.5$ and $-2.61$, within about $2\%$ of the analytic envelope $-\amax(q)$ ($-21.1$ and $-2.56$; Eq.~\eqref{eq26}, Table~\ref{tab01}), a reasonable result for a finite-order fit extrapolated beyond its range.
Figure~\ref{fig02}(a) shows the wells, Fig.~\ref{fig04}(a) the time-domain evolution at $\ell=1$, and Fig.~\ref{fig04}(c) the rates fitted at $\ell=1,2$.
These evolutions test the finite-$\ell$ thresholds, not the envelope. The eikonal argument controls the envelope analytically for all higher multipoles, so the window below is not extrapolated from a finite-$\ell$ scan.
They decay inside the envelope, at it, and just beyond it, and grow only well beyond it. At $\ell=1$, $q=0.8M$ the change from decay to growth falls between the two runs that bracket $\alpha_c(\ell{=}1)$ (Fig.~\ref{fig04}(a)), consistent with thresholds beyond the envelope. The fitted rates change sign across the computed thresholds and, wherever a bound state exists, agree with its $\mathrm{Im}\,\omega$ to better than $0.1\%$ (Fig.~\ref{fig04}(c)).

The positive edge is set by the even-parity potential, Eq.~\eqref{eq08}, whose denominator vanishes where $4\alpha q^2/r^4$ reaches unity. That radius, $r_{\rm pole}=(4\alpha q^2)^{1/4}$, leaves the horizon once $\alpha$ exceeds
\begin{equation}
  \amax(q)\equiv\frac{r_+^4}{4q^2}.
  \label{eq26}
\end{equation}
The two thresholds combine into the symmetric window $|\alpha|<\amax(q)$, with $\amax=96.020$, $21.080$, $7.290$, $2.560$ and $0.821$ at $q=0.2$, $0.4$, $0.6$, $0.8$ and $0.95$ (Table~\ref{tab01}).
Beyond $\amax$ the even-parity potential is negative throughout $r_+<r<r_{\rm pole}$. This well is proportional to $\ell(\ell+1)$, like the odd-parity one, so the same argument makes $+\amax$ the $\ell\to\infty$ bound-state threshold of the even sector (Sec.~\ref{sec04B} gives the growing modes). The well is singular, however: the potential diverges at $r_{\rm pole}$, where the geometric series in Eq.~\eqref{eq23} stops converging. A truncated-domain evolution just outside $\amax$ can show growth or decay depending on whether the truncation margin retains the well.
For $\alpha>\amax$ the pole carries a forced logarithmic branch, and the wave equation ceases to be a regular Sturm--Liouville problem (Sec.~\ref{sec04B}). At $\alpha=\amax$ exactly the pole sits at the horizon, where the zero of $f$ cancels it: the even potential no longer vanishes at the horizon but tends to the finite value $\ell(\ell+1)(r_+-r_-)/(4r_+^3)$, so an ingoing wave arrives there with the reduced wavenumber $k_h^2=\omega^2-V^{\rm even}(r_+)$ rather than with $\omega$ itself.
Just inside the edge, at $\alpha=0.9\,\amax$, the pole lies inside the horizon, the exterior problem is regular, and all four $(q,\ell)$ combinations tested decay with no growth precursor (Fig.~\ref{fig04}(b)).
The ratio $\amax(q)/r_+^2$ equals $24.5,\,5.74,\,2.25,\,1.00,\,0.477$ at $q=0.2,\,0.4,\,0.6,\,0.8,\,0.95$: the wedge and the effective-field-theory ceiling of Sec.~\ref{sec03C} coincide at $q\simeq0.8$, and the ceiling is far more restrictive at small $q$.

The coincidence is exact: $\amax(q)=r_+^2$ gives $(1+\sqrt{1-q^2})^2=4q^2$, whose root is $q=4M/5$. At $q=0.8M$, and only there, the normalized coupling $\tau\equiv\alpha/\amax(q)$ and the ratio $\alpha/r_+^2$ are the same number. We therefore take $q=0.8M$ as the principal charge and run the coupling at both signs, $\alpha=\pm1M^2$ ($\tau=\pm0.39$). The two points differ only in the sign of $\alpha$. Adding and subtracting the two results separates the first-order response, which reverses with the sign of $\alpha$, from the second-order one, which does not. No expansion has to be fitted to a one-sided sweep.
The third benchmark, $q=0.6M$ with $\alpha=+2M^2$ ($\tau=+0.27$), changes the charge at fixed $|\alpha|$; since $\amax(q)$ grows rapidly as $q$ decreases, the same coupling occupies a different fraction of the wedge, separating the $q$ dependence from the $\tau$ dependence.
A fourth in-wedge point, $q=0.8M$ with $\alpha=+2M^2$ ($\tau=+0.78$), is a near-edge control, and an $\alpha=0$, $q=0.4M$ curve as an angular-shape control in Sec.~\ref{sec06}; the validation against the polarization-summed cross section of Ref.~\cite{ref38} is made at $q=0$, that reference's own background.

\subsection{Frobenius analysis of the positive-$\alpha$ edge}\label{sec04B}

Beyond $\amax$ the even-parity potential, Eq.~\eqref{eq08}, has a pole outside the horizon, at $r_{\rm pole}=(4\alpha q^2)^{1/4}$, where the electric factor $1-4\alpha q^2/r^4$ vanishes. Near it the potential behaves as $C/X$ in the tortoise distance $X\equiv\rs-\rs^{\rm pole}$, with residue $C=\ell(\ell+1)/(4r_{\rm pole})$ by direct symbolic computation. The factor $f(r_{\rm pole})$ from the tortoise Jacobian cancels against the $f$-dependence of the residue. $C$ therefore depends on neither $q$ nor $\alpha$ separately, and it is positive for every physical multipole.
The pole is a regular singular point of the equation, with only a simple pole in the potential, and it admits two local solutions: one vanishes linearly at the pole and is analytic there; the other stays finite at the pole but carries a logarithm, $\Psi_2\simeq1+CX\ln|X|$, whose slope diverges as the pole is approached from either side.
The logarithm is forced: the recursion that builds the second solution term by term cannot be continued past its first step unless $C=0$, and Abel's formula, which constructs the second solution from the first, produces the same $CX\ln|X|$ term.
We also test this local prediction by integrating the full (non-truncated) master equation at three points covering static and dynamic ($\omega\neq0$) cases away from the horizon.

The pole is an interior point of the domain, so the question is not the usual one of self-adjointness at the horizon and at infinity but how a solution is continued across a point where the potential diverges.
On either side of the pole the solution is some mixture of the analytic and the logarithmic branch. The mixture on the horizon side is fixed by the boundary data at the horizon, the mixture on the far side by the data at infinity, and nothing in the equation relates the two. Each mixture carries a finite flux by itself, because the equation has no first-derivative term, so the ambiguity comes from the puncture at the pole, not from any divergence of an individual solution.
Only one choice, the analytic branch alone on both sides, keeps the local energy density finite everywhere, and generic scattering data do not make that choice. Continuation across the pole is therefore non-unique unless a physical connection rule is added.
This is what the margin dependence of the truncated-domain evolutions in Sec.~\ref{sec04A} reflects. The discretization picks a connection rule through its treatment of the near-pole region, and a different truncation margin picks a different rule. Refining the grid cannot remove an ambiguity of the continuum problem.
The positive-$\alpha$ edge is thus a regularity boundary of the even-parity master equation as well as a bound-state threshold. Once a connection rule is imposed a self-adjoint extension exists, and Sec.~\ref{sec04C} identifies the one with finite field energy. The negative-$\alpha$ edge is a bound-state threshold of the ordinary, regular kind.

We compute the growing modes on the finite-energy extension.
With the analytic branch imposed on both sides of the pole, no flux crosses it. The annulus $r_+<r<r_{\rm pole}$ then becomes a self-adjoint problem of its own, separate from the exterior, with decay toward the horizon and a vanishing wave function at the pole. The singular well on the horizon side of the pole binds.
We shoot from the pole along the analytic branch and match to a solution that decays toward the horizon. The node count of the zero-energy solution checks the number of bound states independently. Growing modes appear as soon as the edge is crossed.
At $\alpha=1.005\,\amax$, the leading $\ell=2$ mode grows at $M\,\mathrm{Im}\,\omega=0.17$, $0.23$ and $0.39$ for $q=0.6$, $0.8$ and $0.95$, respectively. The leading $\ell=3$ mode has rates $0.59$, $0.70$ and $0.96$, and $\ell=4$ supports two modes at every charge (three at $q=0.95$). For $\ell=1$, the horizon-truncated well has no bound state at $q=0.6$ and $0.8$. At $q=0.95$, it supports only a weakly bound state just beyond the edge ($M\,\mathrm{Im}\,\omega=0.009$ at $1.005\,\amax$, absent by $1.2\,\amax$). This explains why the $\ell=1$ evolution in Sec.~\ref{sec04A} did not detect the instability.
The rates depend weakly on how far the edge is crossed ($0.23$ to $0.17$ for $\ell=2$ at $q=0.8$ between $1.005$ and $2\,\amax$; Fig.~\ref{fig04}(d)), because they are set by the strength $C\propto\ell(\ell+1)/r_{\rm pole}$ of the singular well rather than by its width.
The positive edge is therefore an instability threshold in the same $\ell\to\infty$ sense as the negative edge. Its distinguishing feature is the singular well, whose definition requires the connection condition in Sec.~\ref{sec04C}.

\subsection{The pole in terms of the physical field}\label{sec04C}

A change of variable that is itself singular at the pole can introduce a divergence that the physical field does not have, or hide one that it does. The energy statements above must therefore be made for the physical field, not for the master variable.
The gauge-invariant route of Sec.~\ref{sec02C1} allows this. The three invariants of the even sector, the radial electric field $E_r$, the combination $\Phi$, and $Z$, are algebraic in the master variable $\chi=-r^2P^{tr}$ and its first derivative, with the relations given there, and $\chi$ is itself gauge invariant, so none of them depends on the residual gauge function.
The field energy weights these invariants with the same factors as Eq.~\eqref{eq10}: the radial electric term carries the electric factor $1-4\alpha q^2/r^4$, and the terms in $\Phi$ and $Z$ carry no coupling.
Substituting the two branches of Sec.~\ref{sec04B} then shows how the physical field behaves on each, not only the master variable.
On the analytic branch all three invariants stay finite at the pole and the electric energy density vanishes linearly there.
On the logarithmic branch the master variable stays finite, but the radial electric field is the master variable divided by the electric factor, which vanishes at the pole. $E_r$ therefore diverges as $1/X$, and the electric energy density with it. The field energy in any neighborhood of the pole diverges logarithmically. On that branch $\Phi$ diverges only logarithmically, which is square-integrable, and $Z$ stays finite. The divergence is carried by the physical electric field, not by the choice of variable.
Both branches stay bounded at the pole and are therefore square-integrable, so the operator is in the limit-circle case there: in the Hilbert space $L^2(\dd\rs)$ with its standard inner product, self-adjoint extensions form a one-parameter family on each side, labeled by the connection rule.
Declaring the analytic branch alone admissible on each side picks out one member of that family, the extension whose domain consists of configurations with finite field energy.

\section{Absorption}\label{sec05}

\begin{figure}[t]
  \centering\includegraphics[width=\textwidth]{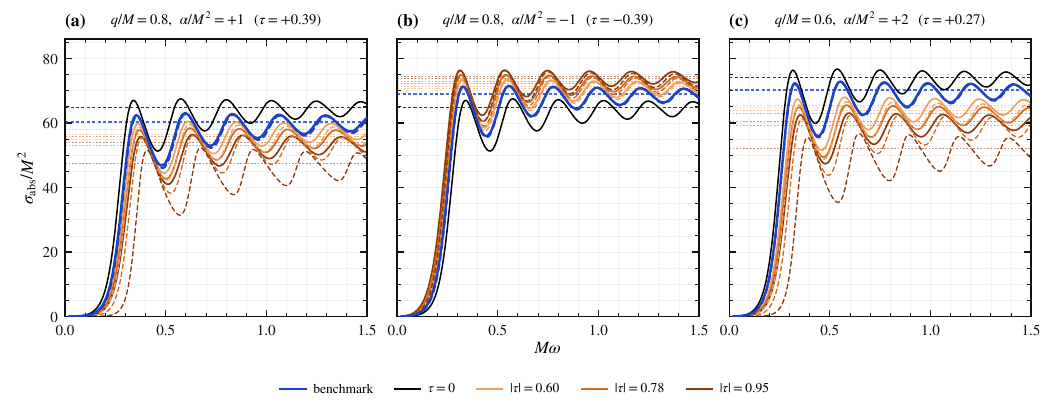}
  \caption{
  Absorption cross section $\sigma_{\rm abs}(\omega)/M^2$ at fixed charge and sign of the coupling, $\tau\equiv\alpha/\amax(q)$.
  Each panel shows the in-wedge benchmark named in its title (blue, thick), the $\tau=0$ Maxwell-on-RN reference (black), and three stronger couplings, $|\tau|=0.60$, $0.78$ and $0.95$ (orange, darkening with $|\tau|$).
  Solid curves are odd parity and dashed curves even; the thin horizontal lines are the geometric-optics limits $\pi(\bc^{\pm})^2$, color-matched to the curves. All panels share one ordinate.}
  \label{fig05}
\end{figure}

\begin{figure}[t]
  \centering\includegraphics[width=\textwidth]{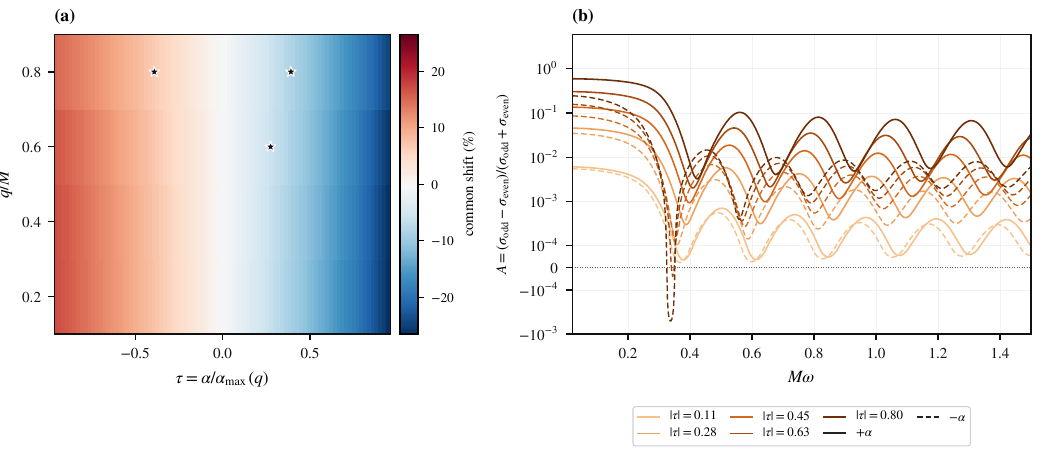}
  \caption{
  Two-order structure of the absorption cross section.
  (a) Parity-common shift $[\tfrac12(\sigma_{\rm abs}^{\rm odd}+\sigma_{\rm abs}^{\rm even})-\sigma_{\rm abs}(\alpha{=}0)]/\sigma_{\rm abs}(\alpha{=}0)$, averaged over $M\omega\geq0.3$, over the $(\tau,q)$ plane.
  (b) Parity asymmetry $A(\omega)$ of Eq.~\eqref{eq28} at $q=0.8$ for $|\tau|=0.11,0.28,0.45,0.63,0.80$, solid for $+\alpha$ and dashed for $-\alpha$, over $0.02\le M\omega\le1.5$ on a symmetric-logarithmic ordinate.}
  \label{fig06}
\end{figure}

We solve Eq.~\eqref{eq04} numerically on an $(\omega,\ell)$ grid with a purely ingoing wave at the horizon.
For the absorption, the solution is matched at a large tortoise radius to an ingoing-plus-outgoing plane-wave basis, giving the gray-body (transmission) factor $\Gamma^{\pm}_\ell$.
For the differential scattering of Sec.~\ref{sec06}, the same radial problem is instead matched to Riccati--Hankel functions, giving the scattering-matrix element $S^{\pm}_\ell$, with $\Gamma^{\pm}_\ell=1-|S^{\pm}_\ell|^2$ linking the two.
The total absorption cross section is
\begin{equation}
  \sigma^{\pm}_{\rm abs}(\omega)=\frac{\pi}{\omega^2}\sum_{\ell\geq1}(2\ell+1)\,\Gamma^{\pm}_\ell(\omega).
  \label{eq27}
\end{equation}
As $\omega\to0$ the cross section vanishes, $\sigma_{\rm abs}\to0$, rather than saturating at a finite monopole value.
This is the standard low-energy behavior of a spin-1 field on a black-hole background, where the absence of a monopole channel forbids the universal low-energy absorption theorem that holds for scalars~\cite{ref39}.
Every curve in Fig.~\ref{fig05} accordingly rises from the origin, staying below $4\times10^{-2}M^2$ at the lowest sampled frequency, $M\omega=0.02$, more than three orders of magnitude under the geometric-optics plateau.
At high frequency both cross sections oscillate about, and converge toward, their geometric-optics limits $\pi(\bc^{\pm})^2$ (Fig.~\ref{fig05}, thin horizontal lines), the standard diffraction pattern set by the unstable photon-sphere orbit and the associated resonances in the complex angular-momentum plane~\cite{ref40}.
Within each panel of Fig.~\ref{fig05} the coupling sequence displaces the whole spectrum, downward for $\alpha>0$ and upward for $\alpha<0$, while the odd and even curves remain nearly coincident at the benchmarks and along the $\alpha<0$ sequence. The splitting becomes visible on this linear scale only toward the positive wedge edge, where $A(M\omega{=}0.3)$ reaches $0.69$ at $q=0.8$, $\tau=0.95$ (Table~\ref{tab02}).

The two-order structure of Sec.~\ref{sec02D} is visible directly in the absorption spectrum.
We define the parity asymmetry
\begin{equation}
  A(\omega)=\frac{\sigma_{\rm abs}^{\rm odd}(\omega)-\sigma_{\rm abs}^{\rm even}(\omega)}{\sigma_{\rm abs}^{\rm odd}(\omega)+\sigma_{\rm abs}^{\rm even}(\omega)},
  \label{eq28}
\end{equation}
which isolates the polarization splitting: it is quadratic in $\alpha$ at leading order, so it survives a reversal of the coupling sign.
The parity-common shift is $\tfrac12(\sigma_{\rm abs}^{\rm odd}+\sigma_{\rm abs}^{\rm even})$ relative to the same-$q$, $\alpha=0$ reference; it isolates the leading, $O(\alpha q^2)$ effect, which is linear in $\alpha$ and therefore reverses with the coupling sign.
Figure~\ref{fig06}(a) shows the parity-common shift over the $(\tau\equiv\alpha/\amax(q),\,q)$ plane: it changes sign with $\alpha$ (positive $\alpha$ contracts the effective cross section, negative $\alpha$ expands it) and spans $-26.5\%$ to $+17.4\%$ on the sampled grid. At fixed normalized coupling $\tau$, its magnitude decreases as $q$ increases because the physical wedge width $\amax(q)$ contracts rapidly with charge; this is the direct, energy-integrated counterpart of the shadow shift reported in Sec.~\ref{sec07}.
Figure~\ref{fig06}(b) overlays $A(\omega)$ for $+\alpha$ and $-\alpha$ at fixed $|\alpha|/\amax$ over the full computed band, $0.02\le M\omega\le1.5$.
The asymmetry is largest at low frequency, where tunneling through the $\ell=1$ barrier exponentially amplifies the small difference between the barriers. It decreases monotonically with $\omega$ toward the eikonal value set by the critical impact parameters. By definition, $A$ is bounded, $|A|\le1$. At $M\omega=0.02$, the radial integration resolves both gray-body factors to a precision much better than their $5$--$10\%$ difference.
Writing $A=a_2\tau^2(1+c\tau+\cdots)$, the ratio $R\equiv A(+\tau)/A(-\tau)$ satisfies exactly $(R-1)/(R+1)=[A(+\tau)-A(-\tau)]/[A(+\tau)+A(-\tau)]$, the part of $A$ antisymmetric under $\alpha\to-\alpha$ relative to its symmetric part, which reduces to $c\tau$ at leading order.
At $M\omega=0.02$, the ratio is $1.02$, $1.11$, $1.31$, $1.47$, $1.58$, $1.94$ and $2.43$ at $|\tau|=0.02$, $0.11$, $0.28$, $0.39$, $0.45$, $0.63$ and $0.80$, respectively. The antisymmetric part is thus $1\%$ of the symmetric part at $|\tau|=0.02$ and $19\%$ at the principal benchmark $|\tau|=0.39$. These values support the quadratic structure for $|\tau|\lesssim0.1$ and to better than $20\%$ at the principal benchmark.
At $M\omega=0.3$, the ratios are $1.09$, $1.57$, $3.26$, $5.25$, $7.07$, $16.8$ and $47.4$. The antisymmetric part is already $68\%$ of the symmetric part at $|\tau|=0.39$. For near-edge couplings $|\tau|\gtrsim0.5$, the expansion no longer controls $A$ at any computed frequency. Growth there reflects the one-sided, pole-driven behavior of $\Weven=(1-x)^{-1}$, not a low-order effect.
At the three principal benchmarks $A(M\omega{=}0.3)=0.0227$, $0.0043$ and $0.0052$, and $A(M\omega{=}0.02)=0.096$, $0.065$ and $0.040$ (Table~\ref{tab02}); the intermediate-frequency values lie below the corresponding parity-common absorption shifts of Fig.~\ref{fig06}(a) and above the geometric rim-splitting values, which probe the potential difference only at the photon sphere.

\begin{figure}[t]
  \centering\includegraphics[width=\textwidth]{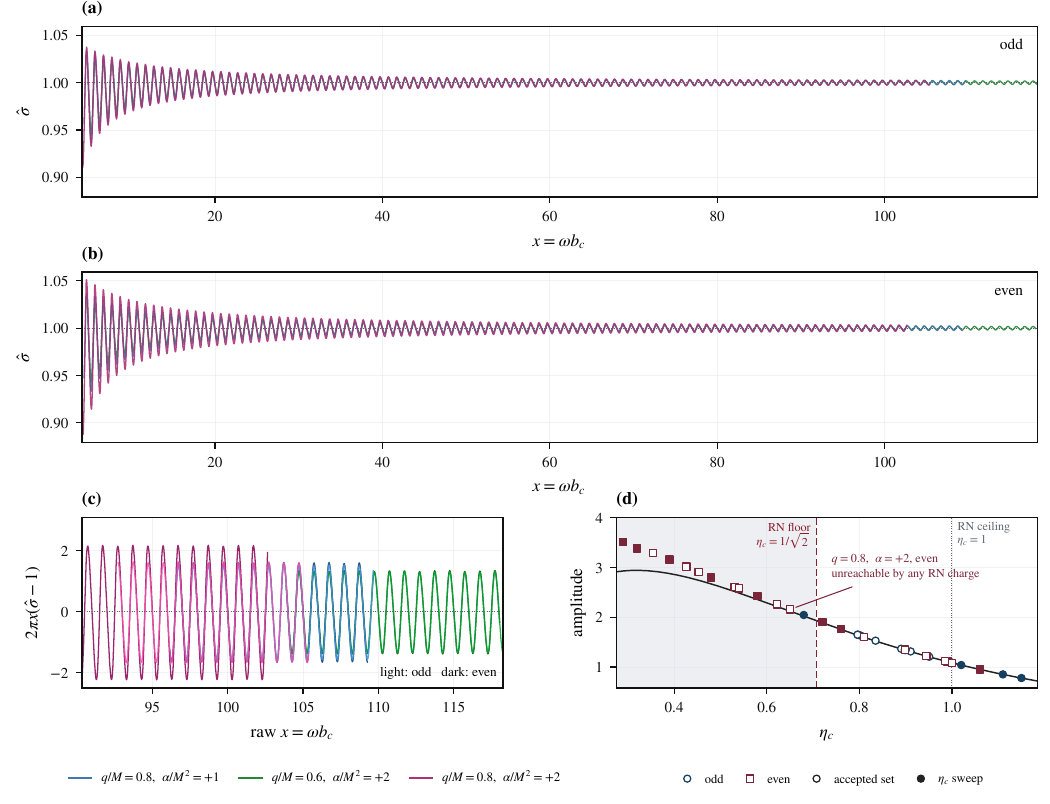}
  \caption{
  Photon-sphere scaling of the absorption fine structure.
  (a),(b) Rescaled cross sections $\hat\sigma^{\pm}=\sigma_{\rm abs}/\pi(\bc^{\pm})^2$ against $x=\omega\bc^{\pm}$, one parity per row and one color per spectrum.
  (c) Amplitude-normalized oscillation $(\hat\sigma^{\pm}-1)\,2\pi\omega/\Omega_c^{\pm}$ in the densely sampled high-frequency windows; hue identifies the spectrum as in (a),(b), lighter for odd and darker for even parity.
  (d) Measured oscillation amplitude against $\eta_c^{\pm}$ for the $27$ spectra of (a),(b) (open symbols) and $12$ points obtained by fixing $\eta_c$ and solving for the coupling (filled symbols), with the eikonal prediction $8\pi\eta_c e^{-\pi\eta_c}$ (solid line). The dotted and dashed lines at $\eta_c=1$ and $1/\sqrt2$ are the upper and lower limits of the Reissner--Nordstr\"om family; the shaded region below the latter contains the even channel of $q=0.8M$, $\alpha=+2M^2$ (labeled).}
  \label{fig07}
\end{figure}

Figure~\ref{fig07} recasts this fine structure as a photon-sphere scaling law, following the data collapse shown for scalar absorption on Frolov black holes in Ref.~\cite{ref23}.
Rescaled by its own photon-sphere pair, each parity's absorption fine structure follows the eikonal form $\hat\sigma^{\pm}-1=-8\pi\eta_c^{\pm}e^{-\pi\eta_c^{\pm}}\,\mathrm{sinc}(2\pi\omega/\Omega_c^{\pm})$~\cite{ref41}, where $\hat\sigma^{\pm}=\sigma^{\pm}_{\rm abs}/\pi(\bc^{\pm})^2$, $\Omega_c^{\pm}=1/\bc^{\pm}$, $\mathrm{sinc}\,z\equiv\sin z/z$, and $\eta_c^{\pm}=\Lambda_c^{\pm}/\Omega_c^{\pm}$ is the ratio of the photon orbit's Lyapunov exponent, $\Lambda_c^{\pm}=f(\rph^\pm)\sqrt{-H^{\pm\prime\prime}(\rph^\pm)/[2H^{\pm}(\rph^\pm)]}$, to its orbital frequency (Table~\ref{tab01}).
The two parities share a common period in $\omega/\Omega_c^{\pm}$ but carry different oscillation amplitudes, set by $\eta_c^{\pm}$ alone.
The measured even/odd amplitude ratio rises from $1.03$ at $q=0.6M$, $\alpha=+2M^2$ to $1.32$ at $q=0.8M$, $\alpha=+2M^2$, and approaches $2$ at the wedge edge.
Rescaling by $\bc^{\pm}$ alone collapses the period but leaves a $130\%$ amplitude spread across our computed spectra; adding the second photon-sphere parameter collapses the amplitudes to $3.5\%$ across all in-wedge spectra, with a worst case of $12.5\%$ at the wedge-edge even point, where the subleading terms of the eikonal expansion are no longer negligible (Fig.~\ref{fig07}).
We performed a dedicated sweep, fixing $\eta_c$ and solving for the required coupling. This adds twelve points at $q=0.4M$--$0.95M$ across both parities, covering $\eta_c\in[0.29,1.15]$ (filled symbols in Fig.~\ref{fig07}(d)).
Over these sampled points, the departure from the leading eikonal amplitude follows a common function of $\eta_c$.
The measured-to-predicted ratio falls smoothly from $1.20$ at $\eta_c=0.29$ through $1.014$ at $\eta_c=0.68$ to within $0.1\%$ of unity for $\eta_c\geq1$.
The odd-parity point at $\eta_c=0.68$, $q=0.95M$ lies between its even-parity neighbors at $q=0.4M$ and $q=0.8M$. This is consistent with parity and charge entering through $\eta_c$ alone over the sampled range.
Ten of the twelve points lie outside $[1/\sqrt2,1]$, the interval accessible to the Reissner--Nordstr\"om family (six below its lower limit and four above its upper limit). They therefore test the scaling beyond the range attainable by varying the ordinary charge.
Over the merged $39$ parameter points the two-parameter spread is $3.7\%$ for $\eta_c\geq0.5$, and the worst case becomes $19.8\%$ at $\eta_c=0.29$, the most extreme near-edge point we compute.

All the trends cited here treat mass $M$ as a constant.
Recasting them at fixed horizon radius $r_+$ reweights the $q$-dependence, since $r_+=M+\sqrt{M^2-q^2}$ relates the two conventions. Reference~\cite{ref23} showed for regular black holes that part of the parameter dependence reported earlier was an artifact of that choice.
The $\alpha$-trends are unchanged by this choice. In a fixed-background test-field treatment, the metric and hence $r_+$ do not depend on $\alpha$, so fixed-$M$ and fixed-$r_+$ presentations of the $\alpha$ dependence coincide.
The principal observables are in any case either dimensionless ratios of same-background quantities (parity asymmetry, rim splitting, $\hat\sigma$) or are plotted against $\omega/\Omega_c^{\pm}$.
Only raw $q$-trends of dimensionful quantities would change presentation at fixed $r_+$, and no conclusion here rests on those.

\section{Scattering and glory}\label{sec06}

\begin{table}
\caption{Combined benchmark results ($M=1$; $q$ in units of $M$, $\alpha$ in units of $M^2$; $\tau\equiv\alpha/\alpha_{\max}(q)$).
The first block lists the three principal in-wedge benchmarks of Figs.~\ref{fig05}--\ref{fig10} and the near-edge control, the second the coupling sequence of Fig.~\ref{fig05}, and the third two further charges at fixed $\tau$.
Ring parameters ($r_{ph}$, $b_c$, rim split) come from the geodesic scan of Sec.~\ref{sec03A}.
$A(0.02)$ and $A(0.3)$ are the absorption parity asymmetry of Eq.~\eqref{eq28} at $M\omega=0.02$ and $0.3$.
$I_{\rm obs}(180^\circ)$ is the polarization-summed backward cross section $|F_-(\pi)|^2/(4\omega^2)$ of Eq.~\eqref{eq29} at $M\omega=4$, one sample of a glory-fringe oscillation of period $0.23$ in $M\omega$, so its size is not comparable between rows.
b.-biref.\ \% is the extremum over $M\omega\in[0.5,4]$ of the per-parity backward split $(I^{\rm odd}_{180}-I^{\rm even}_{180})/\bar I$, with the frequency of the extremum in brackets.}
\label{tab02}

\resizebox{\textwidth}{!}{%
\begin{tabular}{cccccccccccc}
$q$ & $\alpha$ & $\tau$ & $r_{ph}^{\rm odd}$ & $r_{ph}^{\rm even}$ & $b_c^{\rm odd}$ & $b_c^{\rm even}$ & split (\%) & $A(0.02)$ & $A(0.3)$ & $I_{\rm obs}(180^\circ)/M^2$ & b.-biref.\ (\%) [$M\omega$] \\
\hline
\multicolumn{12}{l}{\textit{Principal in-wedge benchmarks}} \\
$0.8$ & $1$ & $+0.39$ & $2.3618$ & $2.3378$ & $4.3860$ & $4.3705$ & $0.341$ & $0.0960$ & $0.02269$ & $0.04505$ & $-6.30$ [$4.00$] \\
$0.8$ & $-1$ & $-0.39$ & $2.6094$ & $2.5970$ & $4.6908$ & $4.6835$ & $0.160$ & $0.0652$ & $0.00429$ & $0.00259$ & $-2.49$ [$3.97$] \\
$0.6$ & $2$ & $+0.27$ & $2.6404$ & $2.6277$ & $4.7301$ & $4.7216$ & $0.174$ & $0.0404$ & $0.00519$ & $0.00161$ & $-2.98$ [$3.93$] \\
$0.8$ & $2$ & $+0.78$ & $2.2540$ & $2.1021$ & $4.2147$ & $4.1057$ & $2.399$ & $0.5616$ & $0.24020$ & $0.39140$ & $-37.8$ [$3.72$] \\
\multicolumn{12}{l}{\textit{Coupling ladder of Fig.~\ref{fig05}}} \\
$0.8$ & $1.548$ & $+0.60$ & $2.3001$ & $2.2281$ & $4.2932$ & $4.2442$ & $1.076$ & $0.2789$ & $0.08836$ & $0.26911$ & $-19.6$ [$3.87$] \\
$0.8$ & $1.990$ & $+0.78$ & $2.2550$ & $2.1056$ & $4.2165$ & $4.1094$ & $2.357$ & $0.5536$ & $0.23480$ & $0.33088$ & $-37.5$ [$3.72$] \\
$0.8$ & $2.432$ & $+0.95$ & $2.2150$ & $1.8828$ & $4.1392$ & $3.8885$ & $5.515$ & $0.9360$ & $0.68564$ & $5.83091$ & $-55.6$ [$1.63$] \\
$0.8$ & $-1.548$ & $-0.60$ & $2.6748$ & $2.6495$ & $4.7637$ & $4.7488$ & $0.326$ & $0.1474$ & $0.00591$ & $0.01486$ & $-4.87$ [$3.91$] \\
$0.8$ & $-1.990$ & $-0.78$ & $2.7257$ & $2.6885$ & $4.8194$ & $4.7977$ & $0.478$ & $0.2340$ & $0.00572$ & $0.09248$ & $-6.92$ [$3.87$] \\
$0.8$ & $-2.432$ & $-0.95$ & $2.7747$ & $2.7250$ & $4.8726$ & $4.8437$ & $0.636$ & $0.3360$ & $0.00422$ & $0.22721$ & $-8.95$ [$3.83$] \\
$0.6$ & $4.407$ & $+0.60$ & $2.5357$ & $2.4511$ & $4.5670$ & $4.5052$ & $1.272$ & $0.2596$ & $0.05785$ & $0.64445$ & $-22.6$ [$3.87$] \\
$0.6$ & $5.666$ & $+0.78$ & $2.4885$ & $2.3158$ & $4.4798$ & $4.3448$ & $2.778$ & $0.5209$ & $0.16531$ & $0.87793$ & $-39.0$ [$3.97$] \\
$0.6$ & $6.926$ & $+0.95$ & $2.4472$ & $2.0739$ & $4.3926$ & $4.0758$ & $6.520$ & $0.9178$ & $0.56959$ & $1.28734$ & $-61.0$ [$1.55$] \\
\multicolumn{12}{l}{\textit{Charge dependence at fixed $\tau$}} \\
$0.2$ & $74.633$ & $+0.78$ & $2.7078$ & $2.5168$ & $4.7416$ & $4.5848$ & $3.038$ & $0.5006$ & $0.10833$ & $2.18369$ & $-39.7$ [$3.34$] \\
$0.4$ & $16.385$ & $+0.78$ & $2.6299$ & $2.4452$ & $4.6474$ & $4.4980$ & $2.956$ & $0.5070$ & $0.12705$ & $4.50917$ & $-39.5$ [$3.41$] \\
\end{tabular}%
}

\end{table}

\begin{figure}[t]
  \centering\includegraphics[width=\textwidth]{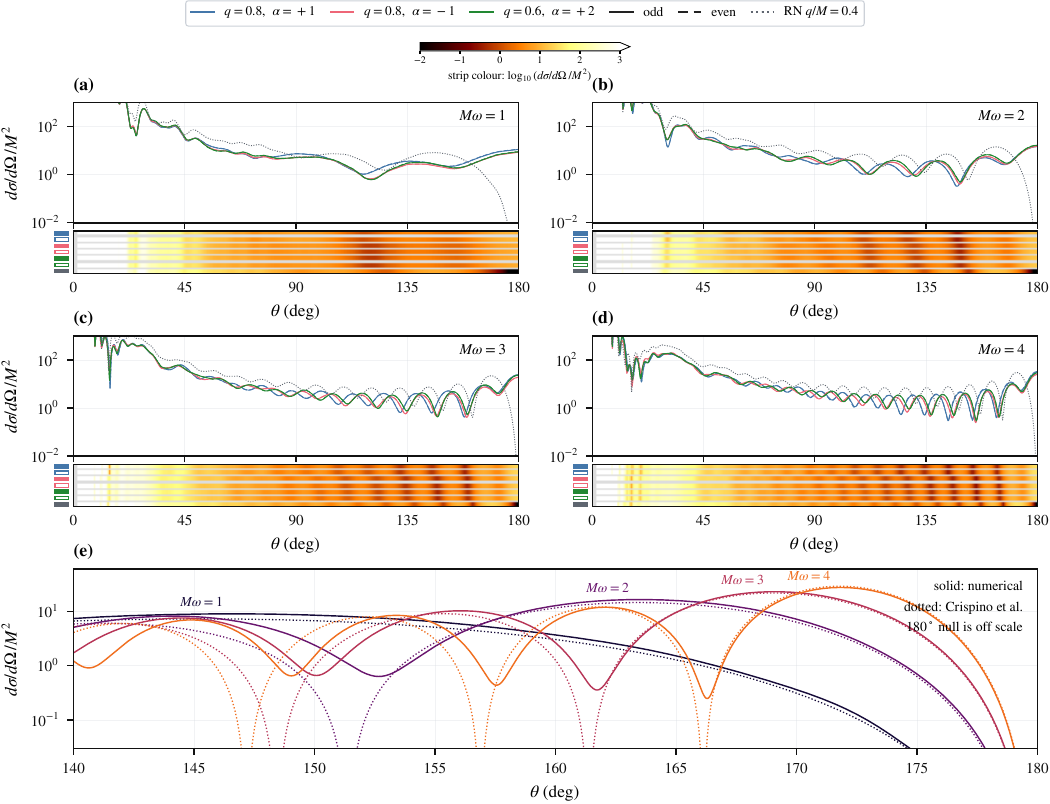}
  \caption{
  (a)--(d) Differential cross section $\dd\sigma/\dd\Omega/M^2$ at $M\omega=1,2,3,4$ for the per-parity decomposition at the three in-wedge benchmarks (odd solid, even dashed) and the combined $\alpha=0$, $q=0.4$ curve (dotted); the ordinate is clipped at $10^{3}$.
  The strip below each panel shows the same seven curves as horizontal bands colored by $\log_{10}$ of the cross section over the same range, in legend order (odd above even for each coupling, then the RN control), with left-edge swatches identifying the coupling (filled odd, open even); the strips start at $\theta=2^\circ$, gray below, and white marks values at or above the upper limit.
  (e) Backward glory of the combined $\alpha=0$ curve at $q=0$ (solid) against the strong-field glory formula of Ref.~\cite{ref38} (dotted).}
  \label{fig08}
\end{figure}

\begin{figure}[t]
  \centering\includegraphics[width=\textwidth]{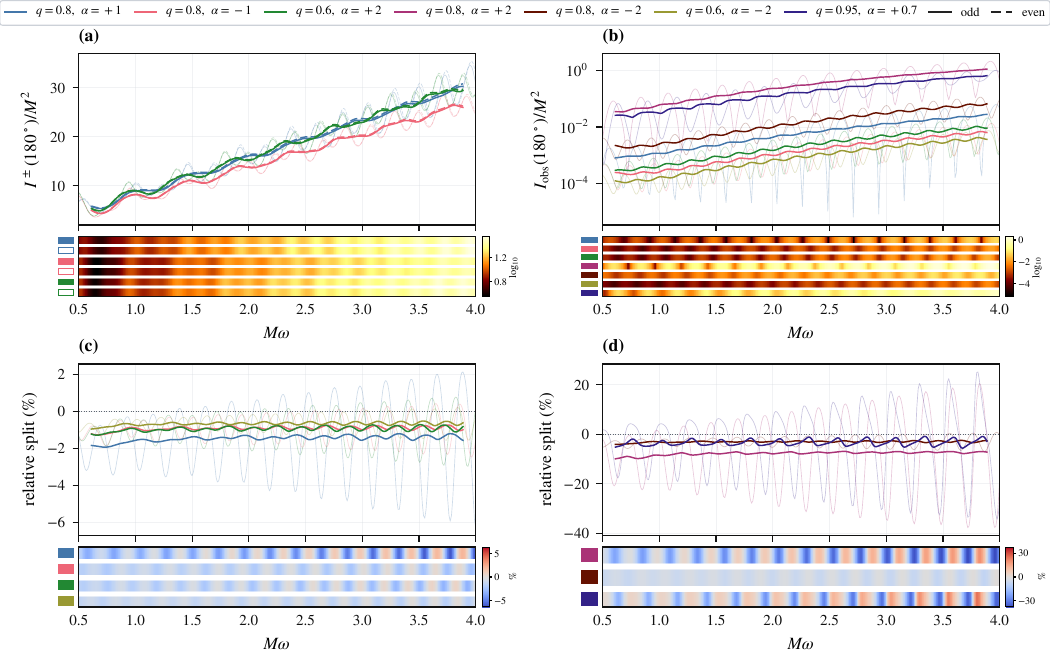}
  \caption{
  Backward scattering. Light traces are the raw point data and solid lines a running mean over one glory-fringe period ($0.23$ in $M\omega$, $21$ samples).
  The strip below each panel repeats that panel's raw curves as horizontal bands, keyed by the left-edge swatches (filled odd, open even in (a)), with the color scale at each strip's right.
  (a) Per-parity decomposition of the backward ($\theta=180^\circ$) intensity at the three in-wedge benchmarks.
  (b) Polarization-summed backward cross section $\dd\sigma/\dd\Omega(180^\circ)=|F_-(\pi)|^2/(4\omega^2)$ of Eq.~\eqref{eq29} at the seven couplings shown in (c) and (d).
  (c) Relative backward split $(I^{\rm odd}_{180}-I^{\rm even}_{180})/\bar I$ of the decomposition at the four moderate couplings, $|\tau|\leq0.39$; the dotted line is the $\alpha=0$ null.
  (d) The same quantity near the wedge edges, $q=0.8M$, $\alpha=\pm2M^2$ and $q=0.95M$, $\alpha=+0.7M^2$, on its own scale.}
  \label{fig09}
\end{figure}

\begin{figure}[t]
  \centering\includegraphics[width=\textwidth]{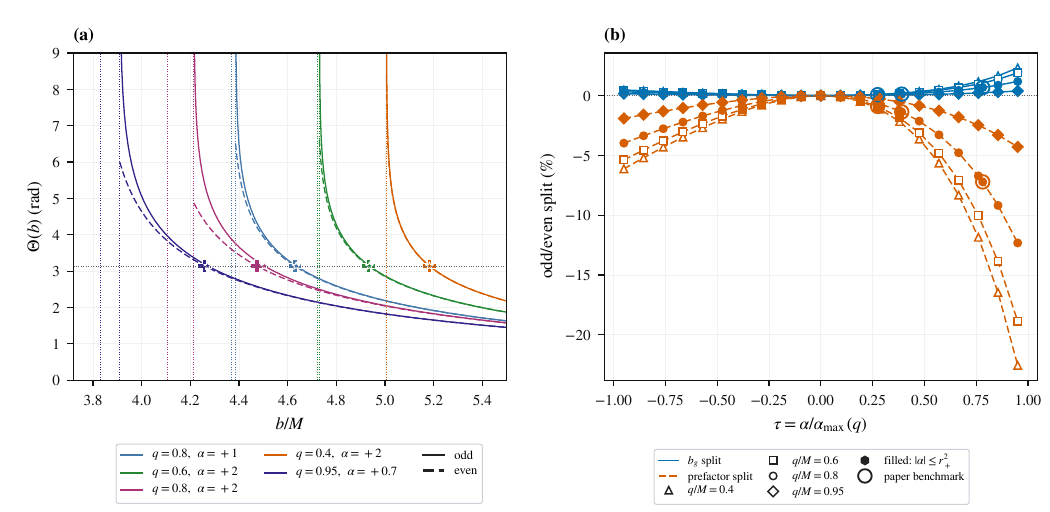}
  \caption{
  Per-parity semiclassical glory.
  (a) Deflection functions $\Theta^{\pm}(b)$ at five couplings from $q=0.4M$ to $0.95M$ (color), odd solid and even dashed; dotted verticals mark $\bc^{\pm}$, and stars (odd) and crosses (even) the glory impact parameters $b_g^{\pm}$ at $\Theta=\pi$.
  (b) Odd/even splits of the glory impact parameter, $(b_g^{\rm odd}-b_g^{\rm even})/\bar b_g$ (blue, solid), and of the glory prefactor, $(A^{\rm odd}-A^{\rm even})/\bar A$ (orange, dashed), against $\tau=\alpha/\amax(q)$ at four charges (marker shape). Filled markers lie inside the effective-field-theory ceiling $|\alpha|\leq r_+^2$ and open markers outside it; large open circles mark the three benchmarks.}
  \label{fig10}
\end{figure}

The differential scattering cross section is built from the same scattering-matrix elements $S^{\pm}_\ell$ using the standard spin-1 partial-wave sum~\cite{ref38,ref42,ref43}.
The field is neutral and has no electromagnetic Coulomb tail. The $2M/r$ gravitational tail nevertheless produces a Coulomb-like, slowly decaying phase at high partial waves, making the sum conditionally convergent. We use a reduced series, multiplying the amplitude twice by $(1-\cos\theta)$ to accelerate convergence~\cite{ref44,ref45}. The sum is truncated at $l_{\max}=\lceil\omega\bc\rceil+40$ (at most $80$), well above the eikonal value $\omega\bc$, with a smooth taper over the highest retained partial waves.
We verified this evaluation by extending the cutoff in steps: every backward-region quantity quoted in this paper is stable to a fraction of a percent, the intermediate-angle fringes to a few percent, and only the forward cone, where the deflection diverges, remains cutoff-limited.
Explicitly, the two parity $S$-matrices enter the observable amplitudes only through their sum and difference,
\begin{equation}
  \begin{aligned}
    F_{\pm}(\theta)&=\sum_{\ell\ge1}\frac{2\ell+1}{\ell(\ell+1)}\,g^{\pm}_{\ell}\,\bigl(\pi_\ell\pm\tau_\ell\bigr),\\
    g^{+}_{\ell}&=\frac{S^{\rm even}_\ell+S^{\rm odd}_\ell}{2}-1,\\
    g^{-}_{\ell}&=\frac{S^{\rm odd}_\ell-S^{\rm even}_\ell}{2},
  \end{aligned}
  \label{eq29}
\end{equation}
with angular functions $\pi_\ell=P^1_\ell/\sin\theta$ and $\tau_\ell=\dd P^1_\ell/\dd\theta$: $F_{+}$ preserves the incident helicity, $F_{-}$ reverses it, and the unpolarized differential cross section is $\dd\sigma/\dd\Omega=(|F_{+}|^2+|F_{-}|^2)/(4\omega^2)$.
The helicity-reversing amplitude is sourced entirely by the parity splitting $S^{\rm odd}_\ell-S^{\rm even}_\ell$, so it vanishes identically at $\alpha=0$; any helicity reversal of a scattered beam in this model is therefore a direct measure of the birefringence.
Figures~\ref{fig08} and~\ref{fig09} present the per-parity cross section and the backward (near $\theta=\pi$) intensity as \emph{per-parity decompositions}: each curve is obtained by setting the \emph{other} channel's scattering matrix to $S_\ell=1$ (flat space) in Eq.~\eqref{eq29}, isolating the contribution of one parity in the calculation.
This is a computational decomposition, not an observable polarization filter. A polarizer cannot isolate ``odd-type-only'' or ``even-type-only'' scattering; a polarization-resolved measurement mixes both channels through the electric- and magnetic-type combinations $S_E,S_M$~\cite{ref46}.
The combined curve at $\alpha=0$ (dotted in Fig.~\ref{fig08}) instead gives the observable, polarization-summed cross section. We use it for comparison, and its $q=0$ counterpart in Fig.~\ref{fig08}(e) for validation against the Schwarzschild electromagnetic scattering calculation of Crispino, Dolan, and Oliveira~\cite{ref38}.
We also compute the unpolarized cross section using both channels' full $S_\ell$ in Eq.~\eqref{eq29} at the benchmarks of Figs.~\ref{fig08} and~\ref{fig09}.
Its helicity-reversal fraction $|F_-|^2/(|F_+|^2+|F_-|^2)$ measures the birefringence directly.
At the principal benchmark ($q=0.8M$, $\alpha=+1M^2$) it stays at the $10^{-5}$--$10^{-4}$ level across $20^\circ$--$160^\circ$ for $M\omega\leq4$.
At the near-edge benchmark ($\alpha=+2M^2$), it reaches a few tenths of a percent at $M\omega=3$--$4$ ($0.9\%$ at $\theta=90^\circ$, $M\omega=3$). At exact backscattering it equals unity wherever $F_-(\pi)\neq0$, which is the generic case at $\alpha\neq0$, because $F_+$ has a backward null. The glory is then entirely helicity-reversing.
The polarization-summed backward intensity, $\dd\sigma/\dd\Omega(180^\circ)=|F_-(\pi)|^2/(4\omega^2)$, is thus sourced entirely by parity splitting. It vanishes identically at $\alpha=0$ for every charge; numerically, we find $10^{-27}M^2$ or less at $q=0.8M$ and $0.6M$ over $0.5\le M\omega\le4$. Figure~\ref{fig09}(b) and Table~\ref{tab02} report this observable over the same band as the per-parity decomposition.
At $M\omega=4$, the intensity is $0.045M^2$ at the principal benchmark, $0.39M^2$ at the near-edge control $q=0.8M$, $\alpha=+2M^2$, and $0.82M^2$ at $q=0.95M$, $\alpha=+0.7M^2$. The per-parity backward glory is about $30M^2$ at the same frequency. Thus between $10^{-3}$ and $3\times10^{-2}$ of either channel's glory survives in the observable helicity-reversed signal, which grows roughly as $\omega^{2}$ across the band.
We quantify the Schwarzschild validation as follows.
At $q=0$ exactly (Crispino \emph{et al.}'s own background) our combined curve is overlaid in Fig.~\ref{fig08}(e) against their analytic strong-field glory formula, $M^{-2}\dd\sigma/\dd\Omega|_{\theta\approx\pi}\simeq30.75\,M\omega\,[J_2(5.36\,M\omega\sin\theta)]^2$ (their Eq.~(13)).
On a $0.05^\circ$ angular grid the glory-peak location agrees to $1.40^\circ$ at $M\omega=1$ and to $0.05^\circ$ at $M\omega\geq2$, and the peak-height agreement improves from $24\%$ at $M\omega=1$ to $5\%$ at $M\omega=3$--$4$, the expected trend for a strong-field formula whose own accuracy improves at higher $M\omega$.
We do not claim to reproduce Crispino, Dolan, Higuchi, and de Oliveira's charged-black-hole backscattered-charge signature~\cite{ref46}.
That effect arises from electromagnetic--gravitational mode conversion: their helicity-reversed amplitude $G$ mixes the electromagnetic and gravitational perturbations through charge-sourced metric backreaction.
It is absent by construction in our fixed-background, test-field treatment, where the vector field never couples to a metric perturbation.
Consistent with this, our combined cross section is exactly isospectral (odd $=$ even, to machine precision) at $\alpha=0$ for \emph{every} $q$, not only $q=0$.
Consequently, each of the polarization-dependent effects reported in this paper requires a coupling constant $\alpha$; in this model, the charge alone does not give rise to any such effects.
Our result is distinct from, and does not reproduce, the charge-alone effect in Ref.~\cite{ref46}.

Figure~\ref{fig08}(a)--(d) shows the per-parity decomposition at $M\omega=1,2,3,4$; the odd and even curves remain nearly coincident across the full angular range at our benchmark couplings, again the wave-optics signature of the reciprocity, and both track the combined $\alpha=0$ curve closely away from small angles.
The ordinate there is clipped at $10^{3}$: the forward diffraction peak reaches $\sim10^{12}$ at $\theta=2^\circ$, and the small-angle region $\theta\lesssim20^\circ$, dominated by the long-range gravitational deflection, is not used quantitatively. The strip maps beneath the panels resolve the large-angle interference minima and their dependence on parity, charge, and coupling.
Figure~\ref{fig08}(e) zooms on the backward glory, the diffractive brightening near $\theta=180^\circ$ produced by rays orbiting near the photon sphere~\cite{ref47,ref48,ref49,ref50}.
The glory peak sharpens and moves toward $180^\circ$ as $M\omega$ grows.
For the neutral, $\alpha=0$ combined curve, the intensity at exactly $\theta=180^\circ$ falls to the electromagnetic backward null (of order $10^{-28}$--$10^{-32}$, far below the plotted range), the same helicity-selection mechanism responsible for the vanishing backward cross section of a spin-1 field in flat and curved backgrounds~\cite{ref51,ref52,ref53}.
Figure~\ref{fig09} shows the per-parity backward intensity (panel a, per-parity decomposition) and the relative backward-birefringence spectrum, $(I^{\rm odd}_{180}-I^{\rm even}_{180})/\bar I$, on two scales (panels c and d).
The split remains at or below $6.3\%$ across $M\omega\in[0.5,4]$ at the four moderate couplings, $|\tau|\leq0.39$ ($q=0.8M$, $\alpha=\pm1M^2$; $q=0.6M$, $\alpha=\pm2M^2$), with fringe-averaged values between $-0.5\%$ and $-2\%$, negative for both signs of $\alpha$, whereas the near-edge control $q=0.8M$, $\alpha=+2M^2$ ($\tau=+0.781$) reaches $-37.8\%$ (fringe average $-7\%$ to $-10\%$).
The extrema quoted here and in Table~\ref{tab02} refer to the raw $313$-point traces of Fig.~\ref{fig09}, not to the fringe-averaged curves.
The $\alpha=0$ null is verified independently on the combined, observable curve.
The two wedge edges produce different backward splits. At matched charge and $|\tau|$, positive coupling gives the larger split throughout the scan, with the disparity increasing with $|\tau|$. The ratio of the maximum $|$split$|$ at $+\alpha$ to that at $-\alpha$ is $1.90$ at $|\tau|=0.274$, $2.54$ at $|\tau|=0.391$, and $5.42$ at $|\tau|=0.781$. At $\alpha=-2M^2$, the split reaches only $7.0\%$, compared with $37.8\%$ at $\alpha=+2M^2$. The split is negative for both signs, consistent with a leading term quadratic in $\alpha$; the pair differs in the sign of $\alpha$.
This is the spectral counterpart of the structural asymmetry established in Sec.~\ref{sec04B}: it is the even-parity factor $(1-x)^{-1}$ that develops a pole as $\alpha\to+\amax$, with no analog on the negative side, where the edge is instead a bound-state threshold.
The near-extremal point $q=0.95M$, $\alpha=+0.7M^2$ ($\tau=+0.852$) reaches $-34.4\%$, comparable to the $q=0.8M$ positive-edge value at a charge $19\%$ larger, so it is proximity to the edge rather than the charge itself that sets the size of the split there.
Table~\ref{tab02} reports the principal-benchmark peaks as $-6.30\%$, $-2.49\%$, and $-2.98\%$ at $M\omega_{\rm peak}=3.93$--$4.00$, together with the near-edge value $-37.8\%$ at $M\omega_{\rm peak}=3.72$.
Across the four sampled charges at fixed $\tau=+0.78$, the rim splitting decreases with charge ($3.04\%$, $2.96\%$, $2.78\%$, $2.36\%$ at $q=0.2$, $0.4$, $0.6$, $0.8$). In contrast, the absorption parity asymmetry at $M\omega=0.3$ increases ($0.108$, $0.127$, $0.165$, $0.235$), while its low-frequency value is nearly charge-independent ($0.50$, $0.51$, $0.52$, $0.55$ at $M\omega=0.02$). These trends reflect different radial weights: the rim split samples the potential difference at the photon sphere, whereas the intermediate-frequency asymmetry probes the barrier region where tunneling ends.
The backward-birefringence band maximum stays nearly charge-independent ($-39.7\%$, $-39.5\%$, $-39.0\%$, $-37.5\%$).

We interpret the backward split through a per-parity extension of the null-geodesic glory analysis in Ref.~\cite{ref23}.
From each optical potential we compute the deflection function $\Theta^{\pm}(b)$, the glory impact parameter $b_g^{\pm}$ solving $\Theta^{\pm}(b_g)=\pi$, and the semiclassical glory prefactor $A^{\pm}=2\pi\omega b_g^2\,|db/d\Theta|_{\Theta=\pi}$; at $q=0$ this procedure reproduces the glory impact parameter $b_g=5.36M$ to $0.06\%$ and the prefactor $30.75$ to $0.008\%$ of the strong-field glory formula quoted above~\cite{ref38}.
As the charge grows the whole deflection structure moves inward, $\bc^{\pm}$ falling from $5.01M$ at $q=0.4M$, $\alpha=+2M^2$ to $3.83$--$3.91M$ at $q=0.95M$, $\alpha=+0.7M^2$ (Fig.~\ref{fig10}(a)).
Across the three benchmarks the parities separate in both quantities, with opposite orderings (Fig.~\ref{fig10}): $b_g^{\rm odd}$ exceeds $b_g^{\rm even}$ by $0.08$--$0.68\%$, while the even channel's glory prefactor exceeds the odd's by $0.9$--$7.2\%$.
The sign of the predicted intensity split agrees with the measured per-parity backward split of Fig.~\ref{fig09}(c) at $62$--$80\%$ of the sampled frequencies, a weak agreement for two oscillating quantities sampled at correlated frequencies. We therefore use the geodesic quantities as a qualitative guide, not as a pointwise prediction.
The sweep over $\tau$ at four charges (Fig.~\ref{fig10}b) shows that the normalized coupling does not by itself fix the split.
At $\tau=+0.174$, the largest value that stays inside the effective-field-theory ceiling $|\alpha|\leq r_+^2$ at every charge sampled, the prefactor split runs $-0.41\%$, $-0.35\%$, $-0.25\%$ and $-0.10\%$ at $q=0.4M$, $0.6M$, $0.8M$ and $0.95M$, a factor of $4.0$ across the charge range.
Both the ordering and a comparable factor ($5.3$ at $|\tau|=0.95$) persist out to the wedge edge, where the low-charge curves have long left that ceiling (open symbols in the figure).
The charge therefore affects this geodesic-level split beyond its role in the normalized coupling. This contrasts with the eikonal absorption amplitude in Sec.~\ref{sec05}, whose departure from the Decanini--Folacci--Raffaelli prediction follows a common function of $\eta_c$ over the sampled points.
The spin-1 glory formula's backward null belongs to the polarization-summed field: $F_+(\pi)$ vanishes kinematically, since $\pi_\ell+\tau_\ell=0$ at $\theta=\pi$ for every $\ell$, and $F_-(\pi)$ vanishes because the odd- and even-parity contributions to the helicity-reversing amplitude cancel when $S^{\rm odd}_\ell=S^{\rm even}_\ell$. The per-parity decomposition removes that cancellation. We therefore interpret the backward split using the geodesic quantities $b_g^{\pm}$ and $A^{\pm}$, not a pointwise $J_2$ comparison.

\section{Shadow and rings}\label{sec07}

\begin{figure}[t]
  \centering
  \includegraphics[width=\textwidth]{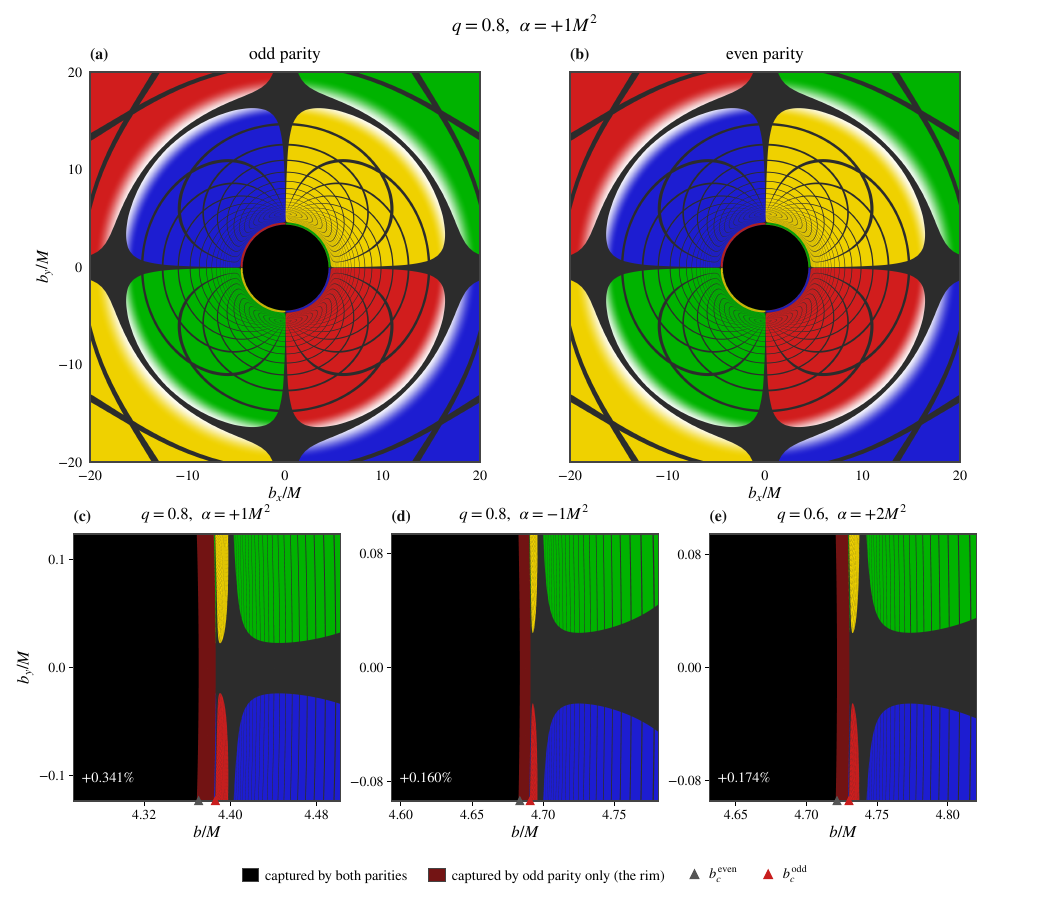}
  \caption{
  Double-shadow images from the two optical metrics, with the observer at $r_{\rm obs}=100M$, the camera calibrated in impact parameter, and a distant four-quadrant celestial sphere carrying a $6^\circ$ coordinate mesh and a $4^\circ$ reference spot on the camera axis.
  (a),(b) Odd- and even-parity shadows at $q=0.8$, $\alpha=+1M^2$ over $|b|\leq20M$.
  (c)--(e) Rim zoom in the impact-parameter plane at $q=0.8,\alpha{=}+1$; $q=0.8,\alpha{=}-1$; and $q=0.6,\alpha{=}+2$. Colors mark the ray's escape quadrant or capture, both parities are traced on the same pixel grid, and the triangles under each axis mark $\bc^{\rm even}$ and $\bc^{\rm odd}$.}
  \label{fig11}
\end{figure}

\begin{figure}[t]
  \centering\includegraphics[width=0.968\textwidth]{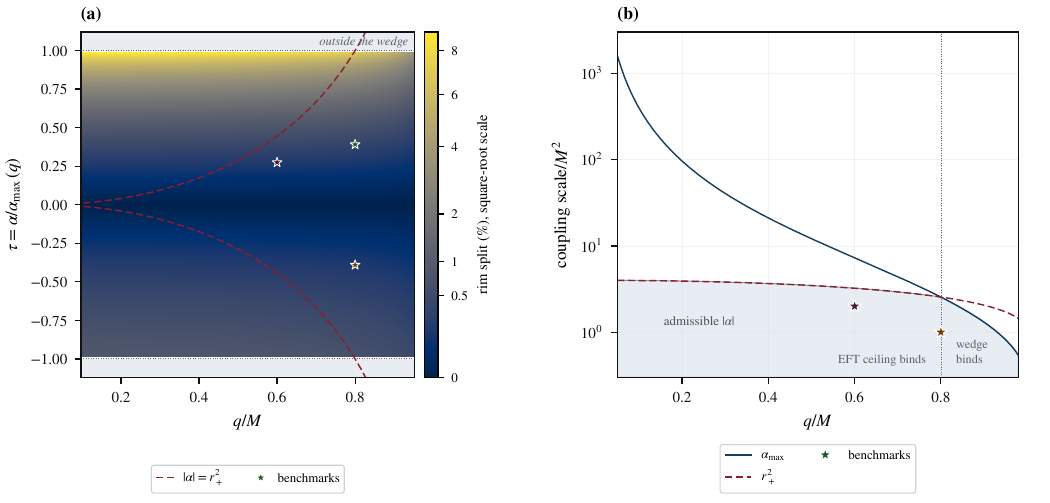}
  \caption{
  (a) Rim splitting $(\bc^{\rm odd}-\bc^{\rm even})/\bc(\alpha{=}0)$, in percent, over the $(\tau,q)$ plane on a square-root color scale, with the effective-field-theory band $|\alpha|=r_+^2$ (dashed) and the three benchmarks (stars); the shaded strips lie outside the wedge $|\tau|<1$.
  (b) Wedge half-width $\amax(q)$ (solid) and effective-field-theory ceiling $r_+(q)^2$ (dashed) on a logarithmic ordinate; the shaded region is the admissible coupling at each charge, the dotted vertical marks their crossing at $q\simeq0.8M$, and stars mark the benchmarks (the $\alpha=\pm1M^2$ pair at $q=0.8M$ coincides).}
  \label{fig12}
  \vspace{-6pt}
\end{figure}

\begin{figure}[!t]
  \centering\includegraphics[width=0.968\textwidth]{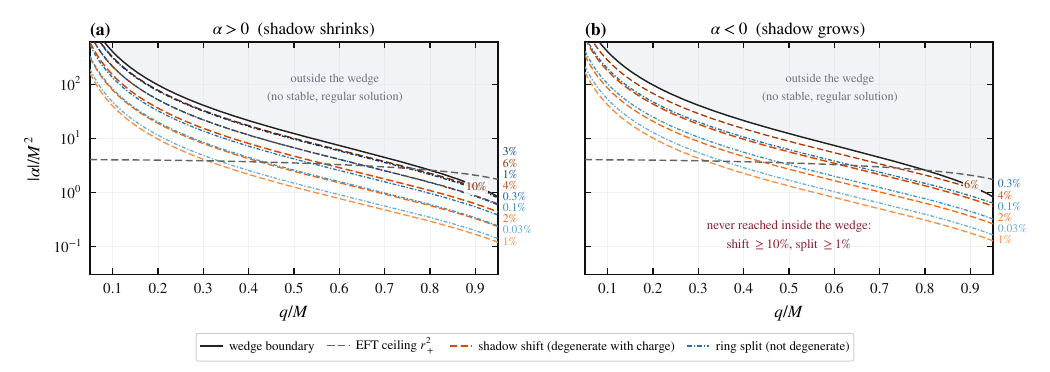}
  \caption{
  Idealized propagation-level sensitivity curves in the $(\alpha,q)$ plane for (a) $\alpha>0$ and (b) $\alpha<0$.
  Each curve marks the coupling at which the labeled observable first reaches the stated level: shift curves (warm, dashed) at $1,2,4,6,10\%$ of the parity-common shadow shift and split curves (cool, dash-dotted) at $0.03,0.1,0.3,1,3\%$ of the rim splitting. Curves absent from (b) correspond to levels not reached inside the wedge.}
  \label{fig13}
  \vspace{-6pt}
\end{figure}

From an observer at $r_{\rm obs}=100M$, with a distant four-color celestial sphere as background~\cite{ref54,ref55}, ray tracing in the optical metrics of Eq.~\eqref{eq24} gives one shadow image for each parity. The images reproduce the analytic $\bc^{\pm}$ in Eq.~\eqref{eq25}, checking consistency between the tracer and wave-optics potentials.
Figures~\ref{fig11}(a)--(b) show the individual-parity images at the principal benchmark, $q=0.8$, $\alpha=+1M^2$. The shadow boundaries are determined by $\bc^{\pm}$, so their difference is confined to the thin annulus between the critical curves, consistent with the $0.34\%$ rim splitting below.
Figures~\ref{fig11}(c)--(e) zoom on that shadow edge for the three in-wedge benchmarks.
The black core, captured in both parities, is bounded by $\bc^{\rm even}$, the smaller critical curve for $\alpha>0$. The narrow colored band between $\bc^{\rm even}$ and $\bc^{\rm odd}$ contains impact parameters captured in one polarization but not in the other.
This band is the double-shadow rim, the image-plane counterpart of the eikonal splitting derived in Sec.~\ref{sec02D}.
Table~\ref{tab01} lists the rim splitting, $(\bc^{\rm odd}-\bc^{\rm even})/\bc(\alpha{=}0)$, and the shadow shift, $\left[\tfrac12(\bc^{\rm odd}+\bc^{\rm even})-\bc(\alpha{=}0)\right]/\bc(\alpha{=}0)$, at the three principal benchmarks: splitting of $0.34\%$, $0.16\%$, and $0.17\%$, always with $\bc^{\rm odd}>\bc^{\rm even}$ for $\alpha>0$, against a shift of $-3.7\%$, $+3.1\%$, and $-2.7\%$.
At a fourth, near-edge point inside the wedge, $q=0.8M$, $\alpha=2M^2$ ($\tau\equiv\alpha/\amax=0.78$, since $\amax=2.56M^2$ there), the shift reaches $-8.5\%$ and the splitting reaches $2.4\%$, a factor of about $3$--$4$ below the shift at that same point.
Both are tabulated in Tables~\ref{tab01} and~\ref{tab02}.
This $2.4\%$ is a benchmark value, not a wedge-wide bound.
Figure~\ref{fig12}(a) maps the rim-splitting percentage over the $(\tau,q)$ plane. It increases toward the positive wedge edge, reaching $8.9\%$ at $\tau=0.98$ for the smallest scanned charges, $q=0.05$--$0.1M$, because $\amax(q)$ grows rapidly as $q\to0$. At $q=0.1M$, this point corresponds to $\alpha\simeq390M^2$, about $100\,r_+^2$, far above the effective-field-theory ceiling. The map stops at $\tau=0.98$, but the splitting remains finite up to the edge: at $\tau=0.9999$ it is $12.2\%$ for $q=0.1M$ and $9.1\%$ for $q=0.8M$, and at $\tau=1$ exactly $12.5\%$ and $9.3\%$. The limit is finite because at the edge $H^{\rm even}=f\,\Weven/r^2$ stays finite at the horizon, $H^{\rm even}(r_+)=f'(r_+)/(4r_+)$; for $q\le\sqrt3M/2$ its maximum sits there, giving $\bc^{\rm even}(\tau{=}1)=r_+^2\sqrt{2/(Mr_+-q^2)}$ and $\eta_c^{\rm even}\to0$, while for larger charge the even photon sphere stays outside the horizon ($r_c^{\rm even}=1.61M$, $\eta_c^{\rm even}=0.27$ at $q=0.95M$). The edge value decreases monotonically with charge, from $12.5\%$ as $q\to0$ to $9.3\%$ at $q=0.8M$ and $3.9\%$ at $q=0.95M$, and on a scan of the whole wedge the splitting never exceeds this $q\to0$ value. In the edge limit the eikonal description deteriorates and, at every charge, the pole reaches the horizon and changes the even-sector ingoing asymptotics (Sec.~\ref{sec04A}).
Away from this narrow near-edge region the splitting stays at a few percent or below, an order of magnitude below the shift at comparable $(\alpha,q)$.
This is the geometric-optics form of the two-order structure found in wave optics (Sec.~\ref{sec05}), and the ray tracer confirms it independently.
Figure~\ref{fig12}(b) shows the wedge half-width $\amax(q)$ itself against the effective-field-theory ceiling $r_+(q)^2$, crossing near $q\simeq0.8$ as noted in Sec.~\ref{sec04A}. In the normalized coordinate of Fig.~\ref{fig12}(a) the ceiling is $\tau_{\rm EFT}=r_+^2/\amax=4q^2/r_+^2$, which reaches unity at $q=4M/5$. The dashed curve leaves the frame there, and beyond that charge the wedge itself is the binding constraint. Below it the ceiling binds and the admissible window is nearly charge-independent, $|\alpha|\lesssim4M^2$; above it the wedge binds and the window closes rapidly, to $|\alpha|\lesssim0.82M^2$ at $q=0.95M$. Across the scanned range $\amax(q)$ falls by three decades, from $1.6\times10^{3}M^2$ at $q=0.05M$ to $0.54M^2$ at $q=0.98M$, which is why comparisons across charge are made in $\tau$ rather than in $\alpha$.
At that crossing charge, restricting to the effective-field-theory band $|\alpha|\lesssim r_+^2$ still allows splitting of about $7\%$ at $\tau=0.98$ and about $9\%$ as $\tau\to1$, so the effective-field-theory ceiling alone does not enforce the $2.4\%$ benchmark value either.
The relevant scales are $0.34\%$ at the principal benchmark and $2.4\%$ at the near-edge control. In the low-charge part of the mathematical wedge, the splitting reaches $8.9\%$ at $\tau=0.98$ and $12.5\%$ as $\tau\to1$, but requires couplings two orders of magnitude above the ceiling.
This shadow phenomenology differs qualitatively from the Weyl-coupled double shadow~\cite{ref04,ref06,ref07,ref08,ref09}, where the splitting itself is the leading, $O(\alpha)$ effect on a vacuum background; here the leading effect is a common shift.
The theoretical basis for interpreting either the shift or the splitting as a signature of new physics rather than an ordinary lensing/mass/spin parameter follows the same photon-ring and critical-curve framework used throughout the shadow literature~\cite{ref56,ref57,ref58,ref59,ref60}.

\enlargethispage{0.8\baselineskip}

\section{Observational reach}\label{sec08}

The field $A_\mu$ is a test field: we do not specify how it is produced near the black hole or how it converts into a signal at a detector, and the images of Sec.~\ref{sec07} are capture maps of the field's two polarizations, not predicted intensity maps for any instrument.
Every curve in this section is therefore an idealized, propagation-level sensitivity forecast. It translates a hypothetical fractional measurement precision on a model observable directly into coupling reach, with the black hole's mass, distance, spin, inclination, and all astrophysical and instrumental nuisance parameters held fixed.
Such forecasts become constraints only when supplemented by a production and detection channel for $A_\mu$ and a full instrument model, both beyond the scope of this paper.
The field $A_\mu$ does not carry the images in Refs.~\cite{ref25,ref27}. The shadow-size precisions below refer to hypothetical measurements of this field's own critical curve, for example through a kinetic-mixing portal to the visible photon with negligible effect on the background. They are not predictions for the visible-light shadow.

The parity-common shadow shift of Sec.~\ref{sec07} is, by construction, degenerate with an ordinary change in the black hole's charge. A shadow-size measurement alone cannot distinguish a shift produced by $\alpha q^2\neq0$ from a shift produced by a different assumed $q$ at $\alpha=0$, because both act on the same critical-curve observable through the same $H(r)$.

Figure~\ref{fig13} makes this limitation explicit. The shift-based curves in $(\alpha,q)$ describe a degeneracy, not independent constraints. The splitting-based curves probe the sub-leading $O(\alpha^2q^4)$ channel, quadratic in $\alpha$, which is not degenerate with charge and is the only internal way to break the degeneracy. An independent measurement of $q$, or a prior on it, is the external alternative.
All sensitivity curves also depend on the assumed charge.
Astrophysical black holes are expected to carry only a small charge, continuously neutralized by ambient plasma, so the shift-based reach below is only as good as an assumed or independently constrained $q\ll M$.
The model itself is regular and stable up to $q=0.95M$; the reach of the shift channel, and its charge degeneracy, are meaningful only relative to this astrophysical charge prior.
Current horizon-scale imaging quotes a fractional shadow-size uncertainty of about $10\%$ for M87*~\cite{ref25} and about $4\%$ for Sgr A*~\cite{ref27}.
The collaboration has already resolved the polarization pattern of the M87* ring~\cite{ref26}, the observational entry point for the polarization-resolved, degeneracy-breaking channel discussed below.
Comparing the total-intensity thresholds to the shift curves of Fig.~\ref{fig13} shows that neither instrument constrains $\alpha$ for astrophysically plausible small charges.
For $\alpha<0$ neither the $10\%$ shift threshold nor a split of $1\%$ or more is reached anywhere inside the stable and regular wedge; the shift saturates below about $9\%$ even at $98\%$ of the wedge half-width.
For $\alpha>0$ it is reached only close to the wedge boundary, where the crossing enters the effective-field-theory band at $q\simeq0.763M$, far from the small charges assumed above.
Breaking the degeneracy with the splitting channel requires photon-ring polarimetry at a level well beyond current Event Horizon Telescope precision, a level targeted by proposed next-generation, space-based very-long-baseline interferometry such as the Black Hole Explorer (BHEX)~\cite{ref61,ref62}.
Its $n=1$ photon-ring polarimetric forecasts for M87* and Sgr A* target fractional precisions at or below the percent level~\cite{ref63,ref64}.
We show representative $1\%$ and $0.1\%$ ring-split thresholds in Fig.~\ref{fig13} as an illustrative target for this channel, consistent with the broader photon-ring interferometric-signature and spacetime-inversion literature~\cite{ref65,ref66,ref67,ref68}.
Under the production and detection assumptions above, near-extremal or dark-charge scenarios with BHEX-class polarimetry are the candidate targets, rather than present-day total-intensity imaging.
We stop the charge sweep at $q=0.95M$. As $q\to1.0M$, the horizon structure and boundary conditions change; the extremal limit is beyond this test-field study.

We next test the degeneracy at the wave level by matching the geometric critical curve as closely as possible.
Reference~\cite{ref23} showed that distinct regular metrics with matched $b_c$ have nearly indistinguishable absorption at intermediate-to-high frequency, and the Kerr--Newman black-bounce family shares its $\bc$ with Kerr--Newman exactly~\cite{ref69}.
Here we ask whether an ordinary Maxwell field on Reissner--Nordstr\"om with a retuned charge $q'$ can mimic the $\alpha\neq0$ spectrum.
For one parity, the agreement is nearly exact: choosing $q'$ such that $b_c^{\rm RN}(q')=b_c^{\rm odd}(q,\alpha)$ reproduces the odd-channel cross section to $0.002$--$0.005\%$ for $\omega M\gtrsim20$, as expected since $\sigma_{\rm abs}\to\pi\bc^2$ there and $\bc$ is matched by construction.
The limitation is that the comparison model has one free parameter, whereas the fine structure depends on two photon-sphere scales.
For the unmatched even channel, the same $q'$ leaves residuals $\delta$, relative to the even-channel cross section, of $0.35\%$, $0.72\%$ and $5.4\%$ at $q=0.6M$, $\alpha=+2M^2$, the principal benchmark $q=0.8M$, $\alpha=+1M^2$, and the near-edge control $q=0.8M$, $\alpha=+2M^2$, respectively. The near-edge control is included because its second photon-sphere scale lies outside the RN range, as discussed below. Equivalently, any $\alpha=0$ comparison model has $A(\omega)\equiv0$, whereas the true high-frequency asymmetry is $\delta/(2+\delta)$, close to half of each residual. These eikonal, $\pi\bc^2$-level mismatches restate the geometric degeneracy at the wave level rather than providing an additional distinction.
The second scale also lies outside the range of this family: across the whole Reissner--Nordstr\"om family $\eta_c=\Lambda_c/\Omega_c$ is bounded below by $1/\sqrt2$, attained at extremality (the closed form $\eta_c(q)$ is given in Ref.~\cite{ref41}), whereas the even channel at $q=0.8M$, $\alpha=+2M^2$ has $\eta_c^{\rm even}=0.650$ (Fig.~\ref{fig07}(d), where the floor and the region below it are marked).
Its eikonal oscillation amplitude, $2.12$ (predicted; $2.16$ measured), exceeds the maximum $1.93$ attainable by any Reissner--Nordstr\"om black hole of any charge.
At that near-edge point no mimic charge in this family reaches the even-parity channel at all; at the three principal benchmarks $\eta_c^{\rm even}=0.810$, $0.945$ and $0.898$ lie inside the RN range, and there the obstruction is the two-scale counting above rather than a hard exclusion.

\section{Conclusions}\label{sec09}

We have studied a neutral vector field coupled to the Einstein tensor on a Reissner--Nordstr\"om background. The coupling vanishes on vacuum spacetimes and is activated by the electric charge.
Two independent derivations give parity potentials that satisfy an exact reciprocity relation.
Compared with the identical-$\alpha$ form, this reciprocity reverses the ordering of the leading and sub-leading effects.
The parity-common shift of the shadow radius, degenerate with an ordinary charge change, is the leading, first-order-in-$\alpha q^2$ effect, reaching $-3.7\%$ at the principal benchmark $q=0.8M$, $\alpha=1M^2$ and $-8.5\%$ at the near-edge point $\alpha=2M^2$.
The birefringent splitting of the two polarizations is a sub-leading effect, second order at small coupling: the rim splitting stays at the few-percent level over most of the stable and regular parameter space and rises only in a thin strip along the positive wedge edge, where it reaches $9.3\%$ at $q=0.8M$ and at most $12.5\%$.
The spectral and angular splitting diagnostics (absorption parity asymmetry, the polarization-summed backward intensity, which vanishes identically at $\alpha=0$, and the per-parity backward split) are small at moderate couplings but grow to the tens-of-percent level close to that edge (Table~\ref{tab02}).
We have also characterized the two edges of the stable and regular window.
Both are bound-state instability thresholds in the $\ell\to\infty$ limit, associated with a potential sign change. The negative-$\alpha$ edge produces a regular odd-parity well, consistent with the finite-multipole thresholds in Ref.~\cite{ref10} and our time-domain evolutions. The positive-$\alpha$ edge produces a singular even-parity well supporting growing modes from $\ell=2$ as soon as the edge is crossed.
A Frobenius analysis shows that the positive edge is also a regularity boundary of the wave equation, with a forced logarithmic singularity, generically divergent local energy density, and a continuation across the pole that is non-unique without an imposed connection condition.
Every effect in this model requires both $\alpha\neq0$ and $q\neq0$, organized into a charge-degenerate leading order and a small sub-leading order that is not degenerate with a charge-only shift: a two-order, charge-activated signature not shared by vacuum-active curvature couplings such as the Weyl coupling.
Breaking the leading-order degeneracy requires either resolving the sub-leading splitting or measuring the black hole's charge independently. The required splitting precision is at the percent level or below, within the range targeted by proposed next-generation photon-ring polarimetry.
Further work could extend the analysis to Kerr--Newman, estimate the coupling from a specific ultraviolet completion, and jointly fit the shift and splitting channels to polarimetric imaging data when a suitable signal becomes available.

\begin{acknowledgments}
This work was supported by the Zhejiang normal university Doctorial research fund Contract No. ZC302924005.
\end{acknowledgments}

\section*{Data availability}
The data are not publicly available. The data are available from the authors upon reasonable request.

\appendix
\begingroup
\section{Two derivations of the parity potentials}
\label{app:parity-derivations}

Both reductions begin with the action in Eq.~\eqref{eq01}. Route I integrates over the vector harmonics before varying the radial action. Route II varies the four-dimensional action first, retains the even-sector gauge function, and eliminates it from the component equations. We give the coupled equations in the original amplitudes $j,h,k,a$, their differential identity, and the intermediate electric-amplitude equation before the final normalization. Keeping $f(r)$ general makes the dependence on its derivatives explicit; RN is imposed only after the two reductions.

\subsection{Conventions and angular identities}
\label{app:derivation-conventions}

Throughout this appendix, the subscripts $\mathrm{odd}$ and $\mathrm{even}$ label the odd- and even-parity sectors, respectively, consistently with the main text.

Write the background as
\begin{equation}
 \begin{aligned}
 g_{\mu\nu}&=\operatorname{diag}
 \left(f,-f^{-1},-r^2,-r^2\sin^2\theta\right),\\
 \sqrt{-g}&=r^2\sin\theta.
 \end{aligned}
 \label{app:metric}
\end{equation}
A prime denotes $\partial_r$, a dot denotes $\partial_{r_*}=f\partial_r$, and time derivatives are written explicitly as $\partial_t$. With the curvature convention of Sec.~\ref{sec01}, direct contraction of the curvature gives
\begin{align}
 g_1\equiv G^t{}_t=G^r{}_r&=\frac{1-f-rf'}{r^2},\nonumber\\
 g_2\equiv G^\theta{}_\theta=G^\phi{}_\phi
 &=-\frac{f''}{2}-\frac{f'}r.
 \label{app:curvature}
\end{align}
The three combinations needed below are
\begin{align}
 \mathcal A&=1-4\alpha g_1
 =1+\frac{4\alpha}{r^2}(rf'+f-1),\nonumber\\
 \mathcal B&=1-4\alpha g_2
 =1+2\alpha f''+\frac{4\alpha f'}r,\nonumber\\
 \mathcal D&=1-2\alpha(g_1+g_2)\nonumber\\
 &=1+\frac{\alpha}{r^2}(r^2f''+4rf'+2f-2).
 \label{app:coefficients}
\end{align}
They are coefficients obtained from the action, not assumed optical factors. The following canonical transformations are made on a connected exterior interval with $f>0$, $\mathcal A\ne0$, and $\mathcal D>0$. The stable RN exterior satisfies these conditions. No transformation through a zero of these coefficients is implied.

Let $\gamma_{ab}$, $D_a$, and $\epsilon_{ab}$ denote the unit-sphere metric, derivative, and area form, with $\epsilon_{\theta\phi}=\sin\theta$. For a real, unit-normalized harmonic $Y=Y_{\ell m}$, define
\begin{equation}
 \begin{aligned}
 D^aD_aY&=-\lambda Y,\\
 X_a&=-\epsilon_a{}^bD_bY,\\
 \lambda&=\ell(\ell+1).
 \end{aligned}
 \label{app:harmonics}
\end{equation}
\nopagebreak[4]
The identities used in the angular integration are
\begin{align}
 &\int Y^2\dd\Omega=1,\quad
 \int D_aY D^aY\dd\Omega=\int X_aX^a\dd\Omega=\lambda,
 \nonumber\\
 &D^aX_a=0,\qquad D_aX_b-D_bX_a=-\lambda\epsilon_{ab}Y.
 \label{app:angular-identities}
\end{align}
Thus the integrated square of the last antisymmetric tensor is $2\lambda^2$. A complex harmonic basis gives the same radial equations with the usual reality condition. We consider $\ell\ge1$: there is no odd monopole, and the even monopole is a constrained Coulomb sector rather than a radiative mode.

In spherical coordinates, the definition of $X_a$ fixes its components as $X_\theta=-\partial_\phi Y/\sin\theta$ and $X_\phi=\sin\theta\,\partial_\theta Y$. For one Fourier mode the full covariant vector potential is therefore

\begin{equation}
 A_\mu=e^{-i\omega t}\left[
 \begin{pmatrix}jY\\hY\\k\partial_\theta Y\\k\partial_\phi Y\end{pmatrix}
 +\begin{pmatrix}0\\0\\-a\,\partial_\phi Y/\sin\theta\\a\sin\theta\,\partial_\theta Y\end{pmatrix}
 \right],\qquad Y=Y_{\ell m}(\theta,\phi).
 \label{app:coordinate-ansatz}
\end{equation}
The two terms have parity $(-1)^\ell$ and $(-1)^{\ell+1}$, respectively. The sign of our odd harmonic is opposite to that of the odd column in Eq.~(4) of Ref.~\cite{ref10}; replacing $a$ by $-a$ relates the conventions and leaves its homogeneous radial equation unchanged. The functions $j,h,k,a$ below refer to this single mode; their $\ell,m$ indices are suppressed.

\subsection{Route I: angular reduction of the action}
\label{app:action-route}

To track the metric signs before the angular integral, define $\mathcal C=G^{\mu\rho}F_{\mu\nu}F_\rho{}^\nu$. Direct contraction gives
\begin{equation}
 \begin{gathered}
 \mathcal C
 =-2g_1F_{tr}^2-\frac{g_1+g_2}{fr^2}\gamma^{ab}F_{ta}F_{tb}
 +\frac{f(g_1+g_2)}{r^2}\gamma^{ab}F_{ra}F_{rb}
 +\frac{g_2}{r^4}\gamma^{ac}\gamma^{bd}F_{ab}F_{cd},\\
 \mathcal L=-\frac14 F_{\mu\nu}F^{\mu\nu}+\alpha\mathcal C
 =\frac{\mathcal A}{2}F_{tr}^2
 +\frac{\mathcal D}{2fr^2}\gamma^{ab}F_{ta}F_{tb}
 -\frac{\mathcal D f}{2r^2}\gamma^{ab}F_{ra}F_{rb}
 -\frac{\mathcal B}{4r^4}\gamma^{ac}\gamma^{bd}F_{ab}F_{cd}.
 \end{gathered}
 \label{app:component-lagrangian}
\end{equation}

Angular indices in these contractions use the unit-sphere metric. Multiplication by $\sqrt{-g}$ and the identities in Eq.~\eqref{app:angular-identities} then give the two radial actions below.

For even parity, set
\begin{equation}
 \begin{aligned}
 A_t&=j(t,r)Y,\\
 A_r&=h(t,r)Y,\\
 A_a&=k(t,r)D_aY.
 \end{aligned}
 \label{app:even-ansatz}
\end{equation}
The nonzero field strengths are
\begin{align}
 F_{tr}&=(\partial_t h-j')Y,\nonumber\\
 F_{ta}&=(\partial_t k-j)D_aY,\nonumber\\
 F_{ra}&=(k'-h)D_aY.
 \label{app:even-field}
\end{align}
The angular integrals entering Eq.~\eqref{app:component-lagrangian} are
\begin{align}
 \int F_{tr}^2\dd\Omega&=(\partial_t h-j')^2,\nonumber\\
 \int\gamma^{ab}F_{ta}F_{tb}\dd\Omega&=\lambda(\partial_t k-j)^2,\nonumber\\
 \int\gamma^{ab}F_{ra}F_{rb}\dd\Omega&=\lambda(k'-h)^2,\nonumber\\
 F_{ab}&=0.
 \label{app:even-angular-contractions}
\end{align}
The factor $r^2$ in $\sqrt{-g}$ then gives the radial action
\begin{align}
 S_{\mathrm{even}}=\frac12\int\dd t\dd r\,\bigg[
 &r^2\mathcal A(\partial_t h-j')^2
 +\frac{\lambda\mathcal D}{f}(\partial_t k-j)^2\nonumber\\
 &-\lambda\mathcal D f(k'-h)^2\bigg].
 \label{app:even-action}
\end{align}
The curvature coupling modifies the radial electric and mixed-component terms through $\mathcal A$ and $\mathcal D$, respectively; their signs are fixed before constraint elimination. To display the constraints before any gauge choice, write
\begin{equation}
 e=\partial_t h-j',\quad v=\partial_t k-j,\quad z=k'-h,\quad p=r^2\mathcal A e.
 \label{app:action-electric-combinations}
\end{equation}
The coefficients of $\delta j$, $\delta h$, and $\delta k$ in the variation of Eq.~\eqref{app:even-action} are, respectively,
\begin{align}
 \mathcal E_j&=p'-\frac{\lambda\mathcal D}{f}v=0,\nonumber\\
 \mathcal E_h&=-\partial_t p+\lambda\mathcal D fz=0,\nonumber\\
 \mathcal E_k&=-\frac{\lambda\mathcal D}{f}\partial_t v
                  +\lambda(\mathcal D fz)'=0.
 \label{app:action-constraints-unfixed}
\end{align}
They obey the identity
\begin{equation}
 -\partial_t\mathcal E_j-\partial_r\mathcal E_h+\mathcal E_k\equiv0.
 \label{app:action-noether-identity}
\end{equation}
Indeed, the two mixed derivatives of $p$ cancel, as do the derivatives of $\lambda\mathcal D v/f$ and $\lambda\mathcal D fz$. Thus the $k$ equation is redundant once the $j$ and $h$ equations hold.

Under $A_\mu\mapsto A_\mu+\partial_\mu(\varepsilon Y)$, the mode functions transform as
\begin{equation}
 j\mapsto j+\partial_t\varepsilon,\quad
 h\mapsto h+\varepsilon',\quad k\mapsto k+\varepsilon.
 \label{app:gauge-transform}
\end{equation}
For this route choose $\varepsilon=-k$, so that $k=0$, while retaining both $j$ and $h$ in the variation. Equation~\eqref{app:action-noether-identity} shows explicitly why the omitted $k$ equation is recovered. The constraint supplied by $j$ is retained.

Introduce an auxiliary momentum $p$ by replacing the electric term with
\begin{equation}
 \frac{r^2\mathcal A}{2}(\partial_t h-j')^2
 \longrightarrow p(\partial_t h-j')-\frac{p^2}{2r^2\mathcal A}.
 \label{app:momentum-introduction}
\end{equation}
Variation of $p$ recovers $p=r^2\mathcal A(\partial_t h-j')$, so the replacement is exact where $\mathcal A\ne0$. Integrating the terms linear in $h$ and $j$ by parts, with compactly supported variations, gives
\begin{align}
 \mathcal L_{\mathrm{even}}^{(1)}={}&-(\partial_t p)h+p'j-\frac{p^2}{2r^2\mathcal A}
 \nonumber\\
 &+\frac{\lambda\mathcal D}{2f}j^2
 -\frac{\lambda\mathcal D f}{2}h^2.
 \label{app:first-order-action}
\end{align}
Varying the remaining potential components gives the algebraic solutions
\begin{equation}
 j=-\frac{fp'}{\lambda\mathcal D},\qquad
 h=-\frac{\partial_t p}{\lambda\mathcal D f}.
 \label{app:auxiliary-components}
\end{equation}
The cancellation of the constrained components can also be seen by completing their squares:

\begin{equation}
 \begin{aligned}
 \mathcal L_{\mathrm{even}}^{(1)}={}&
 -\frac{\lambda\mathcal D f}{2}\left(h+\frac{\partial_t p}{\lambda\mathcal D f}\right)^2
 +\frac{\lambda\mathcal D}{2f}\left(j+\frac{fp'}{\lambda\mathcal D}\right)^2\\
 &+\frac{(\partial_t p)^2}{2\lambda\mathcal D f}
 -\frac{f(p')^2}{2\lambda\mathcal D}-\frac{p^2}{2r^2\mathcal A}.
 \end{aligned}
 \label{app:auxiliary-squares}
\end{equation}

On Eq.~\eqref{app:auxiliary-components}, both squares vanish, leaving
\begin{align}
 S_{\mathrm{even}}=\frac{1}{2\lambda}\int\dd t\dd r\,
 \bigg[\frac{(\partial_t p)^2}{\mathcal D f}
 -\frac{f(p')^2}{\mathcal D}
 -\frac{\lambda p^2}{r^2\mathcal A}\bigg].
 \label{app:momentum-action}
\end{align}
Changing integration coordinate, $\dd r=f\dd r_*$, gives
\begin{equation}
 \begin{aligned}
 S_{\mathrm{even}}&=\frac{1}{2\lambda}\int\frac{\dd t\dd r_*}{\mathcal D}
 \left[(\partial_t p)^2-\dot p^2-U_{\mathrm{even}}p^2\right],\\
 U_{\mathrm{even}}&=\frac{f\lambda\mathcal D}{r^2\mathcal A}.
 \end{aligned}
 \label{app:even-tortoise-action}
\end{equation}
After $p\propto e^{-i\omega t}$, its Euler--Lagrange equation is
\begin{equation}
 \ddot p-\frac{\dot{\mathcal D}}{\mathcal D}\dot p
 +(\omega^2-U_{\mathrm{even}})p=0.
 \label{app:even-momentum-equation}
\end{equation}
For $p=\sqrt{\mathcal D}\,\Psi_{\mathrm{even}}$, the derivative combination in this particular equation becomes

\begin{equation}
 \ddot p-\frac{\dot{\mathcal D}}{\mathcal D}\dot p
 =\sqrt{\mathcal D}\left[
 \ddot\Psi_{\mathrm{even}}
 +\left(\frac{\ddot{\mathcal D}}{2\mathcal D}
       -\frac{3(\dot{\mathcal D})^2}{4\mathcal D^2}\right)\Psi_{\mathrm{even}}\right].
 \label{app:even-normalization-step}
\end{equation}

The terms proportional to $\dot\Psi_{\mathrm{even}}$ cancel. Division by $\sqrt{\mathcal D}$ gives
\begin{equation}
 \begin{gathered}
 \ddot\Psi_{\mathrm{even}}+(\omega^2-V_{\mathrm{even}})\Psi_{\mathrm{even}}=0,\\
 V_{\mathrm{even}}=U_{\mathrm{even}}-\frac{\ddot{\mathcal D}}{2\mathcal D}
 +\frac{3(\dot{\mathcal D})^2}{4\mathcal D^2}.
 \end{gathered}
 \label{app:even-canonical}
\end{equation}
Neither the introduction of $p$ nor its elimination of $j,h$ divided by $\omega$. This route therefore also supplies the static master equation.

For odd parity take $A_t=A_r=0$ and $A_a=a(t,r)X_a$. The field strengths are
\begin{equation}
 \begin{aligned}
 F_{ta}&=(\partial_t a)X_a,\\
 F_{ra}&=a'X_a,\\
 F_{ab}&=-\lambda a\epsilon_{ab}Y.
 \end{aligned}
 \label{app:odd-field}
\end{equation}
For this sector the three nonzero angular contractions are
\begin{align}
 \int\gamma^{ab}F_{ta}F_{tb}\dd\Omega&=\lambda(\partial_t a)^2,\nonumber\\
 \int\gamma^{ab}F_{ra}F_{rb}\dd\Omega&=\lambda(a')^2,\nonumber\\
 \int\gamma^{ac}\gamma^{bd}F_{ab}F_{cd}\dd\Omega&=2\lambda^2a^2.
 \label{app:odd-angular-contractions}
\end{align}
Inserting them into Eq.~\eqref{app:component-lagrangian} gives
\begin{equation}
 \begin{aligned}
 S_{\mathrm{odd}}=\frac\lambda2\int\dd t\dd r\,
 \bigg[&\frac{\mathcal D}{f}(\partial_t a)^2\\
 &-\mathcal D f(a')^2-\frac{\lambda\mathcal B}{r^2}a^2\bigg].
 \end{aligned}
 \label{app:odd-action}
\end{equation}
Here $\mathcal B$ multiplies the angular magnetic energy, whereas $\mathcal D$ multiplies both kinetic terms. Varying $a(t,r)$ first gives
\begin{equation}
 -\frac{\mathcal D}{f}\partial_t^2 a+(\mathcal D f a')'
 -\frac{\lambda\mathcal B}{r^2}a=0.
 \label{app:odd-time-equation}
\end{equation}
Fourier separation, $\partial_t^2 a=-\omega^2a$, then yields
\begin{equation}
 (\mathcal D f a')'
 +\left(\frac{\mathcal D\omega^2}{f}-\frac{\lambda\mathcal B}{r^2}\right)a=0.
 \label{app:odd-radial}
\end{equation}
In tortoise coordinate this is
\begin{equation}
 \ddot a+\frac{\dot{\mathcal D}}{\mathcal D}\dot a
 +(\omega^2-U_{\mathrm{odd}})a=0,\qquad U_{\mathrm{odd}}=\frac{f\lambda\mathcal B}{r^2\mathcal D}.
 \label{app:odd-unnormalized}
\end{equation}
For $a=\Psi_{\mathrm{odd}}/\sqrt{\mathcal D}$, the corresponding combination is

\begin{equation}
 \ddot a+\frac{\dot{\mathcal D}}{\mathcal D}\dot a
 =\frac1{\sqrt{\mathcal D}}\left[
 \ddot\Psi_{\mathrm{odd}}
 +\left(-\frac{\ddot{\mathcal D}}{2\mathcal D}
       +\frac{(\dot{\mathcal D})^2}{4\mathcal D^2}\right)\Psi_{\mathrm{odd}}\right].
 \label{app:odd-normalization-step}
\end{equation}
Multiplication by $\sqrt{\mathcal D}$ gives
\begin{equation}
 \begin{gathered}
 \ddot\Psi_{\mathrm{odd}}+(\omega^2-V_{\mathrm{odd}})\Psi_{\mathrm{odd}}=0,\\
 V_{\mathrm{odd}}=U_{\mathrm{odd}}+\frac{\ddot{\mathcal D}}{2\mathcal D}
 -\frac{(\dot{\mathcal D})^2}{4\mathcal D^2}.
 \end{gathered}
 \label{app:odd-canonical}
\end{equation}

\subsection{Route II: four-dimensional variation with gauge freedom retained}
\label{app:field-route}

Now vary the unreduced action with the background metric and Einstein tensor held fixed. Writing the curvature coupling with both field strengths covariant gives
\begin{equation}
 \delta\mathcal C
 =\left(G^{\mu\rho}F_\rho{}^\nu-G^{\nu\rho}F_\rho{}^\mu\right)\delta F_{\mu\nu}.
 \label{app:curvature-variation-expanded}
\end{equation}
Antisymmetrizing the variational coefficient yields
\begin{align}
 \delta\mathcal L&=-\frac12 P^{\mu\nu}\delta F_{\mu\nu},\nonumber\\
 P^{\mu\nu}&=F^{\mu\nu}-2\alpha G^{\mu\rho}F_\rho{}^\nu
 +2\alpha G^{\nu\rho}F_\rho{}^\mu.
 \label{app:variation}
\end{align}
Integration by parts with compactly supported variations gives $\nabla_\mu P^{\mu\nu}=0$. The diagonal mixed Einstein tensor then implies, without sums over $\mu,\nu$,
\begin{equation}
 P^{\mu\nu}=[1-2\alpha(G^\mu{}_\mu+G^\nu{}_\nu)]F^{\mu\nu}.
 \label{app:constitutive}
\end{equation}
Thus $P^{tr}=\mathcal A F^{tr}$, $P^{ab}=\mathcal B F^{ab}$, and $P^{ta}=\mathcal D F^{ta}$, $P^{ra}=\mathcal D F^{ra}$. This independently recovers the three coefficients in the reduced action.

Substitute the full coordinate ansatz, Eq.~\eqref{app:coordinate-ansatz}, before eliminating any amplitude. Define the radial residuals $\mathcal H_t,\mathcal H_r,\mathcal H_{\mathrm{even}},\mathcal H_{\mathrm{odd}}$ by the actual field components
\begin{equation}
 \begin{aligned}
 \nabla_\mu P^{\mu t}&=\frac{e^{-i\omega t}Y}{r^4f}\mathcal H_t,
 &\nabla_\mu P^{\mu r}&=\frac{e^{-i\omega t}Y}{r^4}\mathcal H_r,\\
 \nabla_\mu P^{\mu a}&=\frac{e^{-i\omega t}}{r^5f}
 \left[-\mathcal H_{\mathrm{even}}D^aY+\mathcal H_{\mathrm{odd}}X^a\right].
 \end{aligned}
 \label{app:component-residual-definition}
\end{equation}
Here $D^aY$ and $X^a$ are raised with the unit-sphere metric. Thus both angular coordinate equations are retained, including for $m\ne0$. The scalar and vector harmonics are independent, so their respective coefficients must vanish. Written entirely in $f$ and the original radial amplitudes, these coefficients are
\small
\begin{align}
 \mathcal H_t={}&r^2f\bigl[r^2+4\alpha(rf'+f-1)\bigr](-j''-i\omega h')
 +r^2f\bigl[2r+4\alpha(rf''+2f')\bigr](-j'-i\omega h)\nonumber\\
 &+\lambda\bigl[r^2+\alpha(r^2f''+4rf'+2f-2)\bigr](j+i\omega k)=0,
 \label{app:raw-t-equation}\\
 \mathcal H_r={}&r^2\omega\bigl[r^2+4\alpha(rf'+f-1)\bigr](\omega h-ij')
 -\lambda f\bigl[r^2+\alpha(r^2f''+4rf'+2f-2)\bigr](h-k')=0,
 \label{app:raw-r-equation}\\
 \mathcal H_{\mathrm{even}}={}&rf^2\bigl[r^2+\alpha(r^2f''+4rf'+2f-2)\bigr](h'-k'')\nonumber\\
 &+f\left\{rf'\bigl[r^2+\alpha(r^2f''+4rf'+2f-2)\bigr]
       +\alpha f(r^3f'''+4r^2f''-2rf'-4f+4)\right\}(h-k')\nonumber\\
 &+i\omega r\bigl[r^2+\alpha(r^2f''+4rf'+2f-2)\bigr](j+i\omega k)=0,
 \label{app:raw-angular-even}\\
 \mathcal H_{\mathrm{odd}}={}&r\bigl[r^2+\alpha(r^2f''+4rf'+2f-2)\bigr](f^2a''+\omega^2a)\nonumber\\
 &+f\left\{rf'\bigl[r^2+\alpha(r^2f''+4rf'+2f-2)\bigr]
       +\alpha f(r^3f'''+4r^2f''-2rf'-4f+4)\right\}a'\nonumber\\
 &-\lambda f\bigl[r+2\alpha(rf''+2f')\bigr]a=0.
 \label{app:raw-angular-odd}
\end{align}
\normalsize
The even residuals are related off shell by
\begin{equation}
 \lambda\mathcal H_{\mathrm{even}}
 =i\omega r\mathcal H_t-rf\mathcal H_r'+2f\mathcal H_r.
 \label{app:raw-residual-identity}
\end{equation}
Differentiating $\mathcal H_r$ generates the same $j''$ and $h'$ terms as $i\omega r\mathcal H_t$. Their cancellation leaves exactly the $h-k'$ and $j+i\omega k$ terms in $\lambda\mathcal H_{\mathrm{even}}$. Consequently, the two scalar equations determine the even sector, and $\mathcal H_{\mathrm{odd}}=0$ determines the odd sector. This displays the component and elimination steps corresponding to the Appendix of Ref.~\cite{ref10}, with all coefficients derived using the action and curvature convention of this paper.

For the even sector, retain $k$ and define
\begin{align}
 \Phi&=j+i\omega k,\quad Z=h-k',\quad E=-i\omega h-j',\nonumber\\
 E&=-(\Phi'+i\omega Z),\qquad
 \chi=-r^2P^{tr}/Y=r^2\mathcal A E.
 \label{app:invariants}
\end{align}
Here $P^{tr}/Y$ denotes the radial coefficient after the common time factor has also been removed, as in the main text. The field strengths and their raised constitutive components are
\begin{align}
 F_{tr}&=EY,& P^{tr}&=-\chi Y/r^2,\nonumber\\
 F_{ta}&=-\Phi D_aY,& P^{ta}&=\frac{\mathcal D\Phi}{fr^2}D^aY,\nonumber\\
 F_{ra}&=-Z D_aY,& P^{ra}&=-\frac{\mathcal D fZ}{r^2}D^aY.
 \label{app:even-P}
\end{align}
The $t,r$, and angular components, respectively, reduce to
\begin{align}
 \chi'+\frac{\lambda\mathcal D}{f}\Phi&=0,\nonumber\\
 i\omega\chi-\lambda\mathcal D fZ&=0,\nonumber\\
 (\mathcal D fZ)'+\frac{i\omega\mathcal D}{f}\Phi&=0.
 \label{app:even-constraints}
\end{align}
The last equation has an immaterial overall minus sign relative to the angular component of the divergence.
\nopagebreak[4]
Eliminate the first two equations as
\begin{equation}
 \Phi=-\frac{f\chi'}{\lambda\mathcal D},\qquad
 Z=\frac{i\omega\chi}{\lambda\mathcal D f}.
 \label{app:constraint-solution}
\end{equation}

For $\omega\ne0$, the first relation explicitly eliminates the gauge amplitude:
\begin{equation}
 k=\frac{i}{\omega}\left(j+\frac{f\chi'}{\lambda\mathcal D}\right),\qquad
 h-k'=\frac{i\omega\chi}{\lambda\mathcal D f}.
 \label{app:explicit-k-elimination}
\end{equation}
Equivalently, for an arbitrary choice of $k$, the reconstructed potentials are $j=-f\chi'/(\lambda\mathcal D)-i\omega k$ and $h=k'+i\omega\chi/(\lambda\mathcal D f)$. Inserting these in the electric amplitude shows the cancellation directly:
\begin{equation}
 \begin{aligned}
 -i\omega h-j'
 &=-i\omega\left(k'+\frac{i\omega\chi}{\lambda\mathcal D f}\right)
   +\left(\frac{f\chi'}{\lambda\mathcal D}\right)'+i\omega k'\\
 &=\frac1\lambda\left(\frac f{\mathcal D}\chi'\right)'
   +\frac{\omega^2\chi}{\lambda\mathcal D f}.
 \end{aligned}
 \label{app:electric-gauge-cancellation}
\end{equation}
Only the explicit solution for $k$ in Eq.~\eqref{app:explicit-k-elimination} divides by $\omega$. The reconstruction with arbitrary $k$ and the invariant equations below also apply at $\omega=0$.
\begin{samepage}
These expressions satisfy the angular equation identically, including at $\omega=0$. Substitution into the Bianchi relation in Eq.~\eqref{app:invariants} gives
\begin{align}
 \frac{\chi}{r^2\mathcal A}
 &=\frac1\lambda\left(\frac f{\mathcal D}\chi'\right)'
 +\frac{\omega^2\chi}{\lambda\mathcal D f}.
 \label{app:bianchi-elimination}\\
 \intertext{Multiplying by $\lambda\mathcal D f$ and expanding,}
 f^2\chi''+
 \left(ff'-f^2\frac{\mathcal D'}{\mathcal D}\right)\chi'
 &+\left(\omega^2-\frac{f\lambda\mathcal D}{r^2\mathcal A}\right)\chi=0.
 \label{app:chi-radial}
\end{align}
\end{samepage}
Since $\ddot\chi=f^2\chi''+ff'\chi'$, this agrees with Eq.~\eqref{app:even-momentum-equation} under $\chi=p$. The same normalization $\Psi_{\mathrm{even}}=\chi/\sqrt{\mathcal D}$ gives Eq.~\eqref{app:even-canonical}. The equality $p=\chi$ is established after the two reductions, rather than imposed to select a potential.

For odd parity, keep the time-dependent amplitude $a(t,r)$ in Eq.~\eqref{app:odd-field}. Raising the indices and applying the constitutive relation gives
\begin{equation}
 P^{ta}=-\frac{\mathcal D}{fr^2}(\partial_t a)X^a,\qquad
 P^{ra}=\frac{\mathcal D f}{r^2}a'X^a,\qquad
 P^{ba}=-\frac{\lambda\mathcal B}{r^4}a\epsilon^{ba}Y,\qquad P^{tr}=0.
 \label{app:odd-constitutive-components}
\end{equation}
\begin{samepage}
Using $\epsilon^{ba}D_bY=X^a$ and $D_aX^a=0$, the component divergences are
\begin{equation}
 \begin{aligned}
 \nabla_\mu P^{\mu t}
 &=\frac{\mathcal D}{fr^2}(\partial_t a)D_bX^b=0,
 &\nabla_\mu P^{\mu r}
 &=-\frac{\mathcal D f}{r^2}a'D_bX^b=0,\\
 D_bP^{ba}&=-\frac{\lambda\mathcal B}{r^4}aX^a,
 &\nabla_\mu P^{\mu a}
 &=\frac{X^a}{r^2}\left[-\frac{\mathcal D}{f}\partial_t^2 a
 +(\mathcal D f a')'-\frac{\lambda\mathcal B}{r^2}a\right].
 \end{aligned}
 \label{app:odd-divergence-expanded}
\end{equation}
\end{samepage}
Here $D_bP^{ba}$ is the unit-sphere covariant divergence at fixed $t,r$. After $a(t,r)=a(r)e^{-i\omega t}$, the coefficient of $X^a$ is exactly Eq.~\eqref{app:odd-radial}. The rescaling $\Psi_{\mathrm{odd}}=\sqrt{\mathcal D}\,a$ then yields Eq.~\eqref{app:odd-canonical} without using the reduced action.

To track the intermediate electric variable in the component elimination, introduce
\begin{equation}
 Q=\frac{r^2}{\lambda}(-i\omega h-j'),\qquad
 \chi=\lambda\mathcal A Q.
 \label{app:raw-Q-definition}
\end{equation}
This $Q$ has the same definition as Eq.~(A6) of Ref.~\cite{ref10}. It is gauge invariant, but it is not yet the canonical amplitude. Substituting $\chi=\lambda\mathcal A Q$ into Eq.~\eqref{app:chi-radial} gives the equation before normalization:
\begin{equation}
 Q''+\left(\frac{f'}f+\frac{2\mathcal A'}{\mathcal A}
                    -\frac{\mathcal D'}{\mathcal D}\right)Q'
 +\left[\frac{\omega^2}{f^2}-\frac{\lambda\mathcal D}{r^2f\mathcal A}
       +\frac{\mathcal A''}{\mathcal A}
       +\left(\frac{f'}f-\frac{\mathcal D'}{\mathcal D}\right)
        \frac{\mathcal A'}{\mathcal A}\right]Q=0.
 \label{app:raw-Q-equation}
\end{equation}
For explicit background coefficients define the following polynomials, distinct from the coupling contraction $\mathcal C$:
\begin{align}
 C(r)&=r^2\mathcal A=r^2+4\alpha(rf'+f-1),
 &T(r)&=r^2\mathcal D=r^2+\alpha S(r),\nonumber\\
 R(r)&=r\mathcal B=r+2\alpha(rf''+2f'),
 &S(r)&=r^2f''+4rf'+2f-2.
 \label{app:background-polynomials}
\end{align}
The two unnormalized radial equations can then be written as
\begin{align}
 a''+\frac{\dd}{\dd r}\ln\left|\frac{fT}{r^2}\right|a'
 +\left(\frac{\omega^2}{f^2}-\frac{\lambda R}{rfT}\right)a&=0,
 \label{app:odd-log-equation}\\
 Q''+\frac{\dd}{\dd r}\ln\left|\frac{fC^2}{r^2T}\right|Q'
 +\left[\frac{\omega^2}{f^2}-\frac{\lambda T}{r^2fC}
 +\frac{\alpha N_1+\alpha^2N_2}{r^2fCT}\right]Q&=0,
 \label{app:even-log-equation}
\end{align}
where the remaining coefficients are
\begin{align}
 N_1={}&4r^2\left\{f(r^3f'''-r^2f''-2rf'+6f-6)
                       +rf'(r^2f''-2f+2)\right\},
 \label{app:Q-polynomial-one}\\
 N_2={}&4\left\{(r^2f''+4rf'+2f-2)
 \left[f(r^3f'''-r^2f''-2rf'+6f-6)+rf'(r^2f''-2f+2)\right]\right.\nonumber\\
 &\left.\hspace{1.5em}-f(r^3f'''+4r^2f''-2rf'-4f+4)(r^2f''-2f+2)\right\}.
 \label{app:Q-polynomial-two}
\end{align}
These terms come from differentiating $\mathcal A Q$. For example,
\begin{equation}
 \mathcal A'=\frac{4\alpha}{r^3}(r^2f''-2f+2),\qquad
 \mathcal A''=\frac{4\alpha}{r^4}(r^3f'''-r^2f''-2rf'+6f-6),
 \label{app:A-background-derivatives}
\end{equation}
which, together with $\mathcal D'/\mathcal D=T'/T-2/r$, reproduces $N_1$ and $N_2$ in Eq.~\eqref{app:raw-Q-equation}.

The required normalizations for Eqs.~\eqref{app:odd-log-equation} and~\eqref{app:even-log-equation} are
\begin{equation}
 a=\frac{r}{\sqrt T}\Psi_{\mathrm{odd}},\qquad
 Q=\frac{r\sqrt T}{C}\widehat\Psi_{\mathrm{even}},\qquad
 \widehat\Psi_{\mathrm{even}}\equiv\frac{\Psi_{\mathrm{even}}}{\lambda}.
 \label{app:component-normalizations}
\end{equation}
The constant $\lambda$ affects only amplitude normalization. For these radial rescaling factors, the first-derivative coefficients satisfy
\begin{equation}
 \begin{aligned}
 2\frac{\dd}{\dd r}\ln\left(\frac r{\sqrt T}\right)
 +\frac{\dd}{\dd r}\ln\left|\frac{fT}{r^2}\right|&=\frac{f'}f,\\
 2\frac{\dd}{\dd r}\ln\left|\frac{r\sqrt T}{C}\right|
 +\frac{\dd}{\dd r}\ln\left|\frac{fC^2}{r^2T}\right|&=\frac{f'}f.
 \end{aligned}
 \label{app:component-first-derivative-cancellation}
\end{equation}
Multiplication by $f^2$ therefore combines the radial derivative terms into $\partial_{r_*}^2$. The zeroth-order terms give the potentials in Eqs.~\eqref{app:even-canonical} and~\eqref{app:odd-canonical}.

Their dependence on the background derivatives can also be displayed without tortoise derivatives of $\mathcal D$. Define
\begin{align}
 J(r)&=r^3f'''+4r^2f''-2rf'-4f+4,\nonumber\\
 K(r)&=r^4f''''+4r^3f'''-6r^2f''+12f-12.
 \label{app:derivative-polynomials}
\end{align}
Then $\mathcal D'=\alpha J/r^3$ and $\mathcal D''=\alpha K/r^4$, so that
\begin{equation}
 \dot{\mathcal D}=\frac{\alpha fJ}{r^3},\qquad
 \ddot{\mathcal D}=\frac{\alpha f}{r^4}(rf'J+fK).
 \label{app:D-background-derivatives}
\end{equation}
Substitution yields the general-background potentials
\begin{align}
 V_{\mathrm{odd}}&=\frac{f\lambda R}{rT}
 +\frac{\alpha f(rf'J+fK)}{2r^2T}
 -\frac{\alpha^2f^2J^2}{4r^2T^2},\nonumber\\
 V_{\mathrm{even}}&=\frac{f\lambda T}{r^2C}
 -\frac{\alpha f(rf'J+fK)}{2r^2T}
 +\frac{3\alpha^2f^2J^2}{4r^2T^2}.
 \label{app:explicit-general-potentials}
\end{align}
Equations~\eqref{app:background-polynomials} and~\eqref{app:derivative-polynomials} specify every coefficient through $f''''$. These higher background derivatives enter when a radially varying kinetic coefficient is normalized; the dynamical equations remain second order in the perturbation. Both routes give these expressions before any RN identity is used.

\subsection{RN specialization and the radial-potential variable}
\label{app:rn-specialization}

Only now set $f=1-2M/r+q^2/r^2$. The background derivatives are
\begin{align}
 f'&=\frac{2M}{r^2}-\frac{2q^2}{r^3},&
 f''&=-\frac{4M}{r^3}+\frac{6q^2}{r^4},\nonumber\\
 f'''&=\frac{12M}{r^4}-\frac{24q^2}{r^5},&
 f''''&=-\frac{48M}{r^5}+\frac{120q^2}{r^6}.
 \label{app:RN-background-derivatives}
\end{align}
They give $S=J=K=0$, $T=r^2$, $C=r^2-4\alpha q^2/r^2$, and $R=r+4\alpha q^2/r^3$. Equivalently,
\begin{equation}
 \mathcal A=1-\frac{4\alpha q^2}{r^4},\quad
 \mathcal B=1+\frac{4\alpha q^2}{r^4},\quad \mathcal D=1.
 \label{app:rn-factors}
\end{equation}
Both derivative corrections in Eqs.~\eqref{app:even-canonical} and~\eqref{app:odd-canonical} therefore vanish. With $\Psi_{\mathrm{even}}=p=\chi$ and $\Psi_{\mathrm{odd}}=a$, both routes give
\begin{align}
 V_{\mathrm{odd}}&=\frac{f\lambda}{r^2}\left(1+\frac{4\alpha q^2}{r^4}\right),\nonumber\\
 V_{\mathrm{even}}&=\frac{f\lambda}{r^2(1-4\alpha q^2/r^4)}.
 \label{app:rn-potentials}
\end{align}

The intermediate equation for the radial potential component can also be displayed explicitly. In the $k=0$ gauge and for $\omega\ne0$, Eq.~\eqref{app:auxiliary-components} gives
\begin{equation}
 h=\frac{i\omega p}{\lambda f},\qquad
 j=-\frac{f}{i\omega}(fh)'.
 \label{app:h-map}
\end{equation}
Using $E=-i\omega h-j'$ and $p=r^2\mathcal A E$ then gives
\begin{equation}
 [f(fh)']'+\left(\omega^2-\frac{f\lambda}{r^2\mathcal A}\right)h=0,
 \label{app:h-equation}
\end{equation}
or, on expanding the derivatives,
\begin{equation}
 \begin{aligned}
 &f^2h''+3ff'h'\\
 &\qquad+[\omega^2-V_{\mathrm{even}}+ff''+(f')^2]h=0.
 \end{aligned}
 \label{app:h-expanded}
\end{equation}
The terms $ff''+(f')^2$ are independent of $\alpha$, but this equation is not a Schr\"odinger normal form: in tortoise coordinate it also contains $2f'\dot h$. Set $u=fh=i\omega p/\lambda$ and denote the left-hand side of Eq.~\eqref{app:h-expanded} by $\mathcal H[h]$. Then

\begin{equation}
 \begin{aligned}
 \mathcal H[u/f]
 &=fu''+f'u'+\frac{\omega^2-V_{\mathrm{even}}}{f}u
 =\frac1f\left[\ddot u+(\omega^2-V_{\mathrm{even}})u\right]=0.
 \end{aligned}
 \label{app:h-cancellation}
\end{equation}

Thus the geometric terms cancel, and the remaining radial derivatives combine into $\ddot u$. They are not an additional physical potential. At $\omega=0$, $h$ cannot parametrize a nonzero static electric momentum through this map; Eqs.~\eqref{app:momentum-action} and~\eqref{app:bianchi-elimination} give the static equation directly. Likewise, a raw electric amplitude $E$ requires the radial normalization $\Psi_{\mathrm{even}}=r^2\mathcal A E$, not merely a constant multiple of $E$.

\subsection{Relation to the published master variable}
\label{app:published-variable}

The intermediate $Q$ in Eq.~\eqref{app:raw-Q-definition} agrees with the definition in Ref.~\cite{ref10}. However, its multiplier in the canonical variable here is $\mathcal A/\sqrt{\mathcal D}$, as Eq.~\eqref{app:component-normalizations} shows. This should be distinguished from their printed even master variable. On RN, their Eq.~(8) reduces to
\begin{equation}
 \Psi_{\rm pub}=\frac{r^2E}{\lambda(1+x)},\qquad
 x=\frac{4\alpha q^2}{r^4}.
 \label{app:published-variable-rn}
\end{equation}
Our momentum instead satisfies
\begin{equation}
 \chi=\lambda(1-x^2)\Psi_{\rm pub}.
 \label{app:variable-ratio}
\end{equation}
This is a nonconstant radial factor for $\alpha q\ne0$. To see its consequence without choosing a potential by inspection, put $\mathcal K=1-x^2$ and substitute Eq.~\eqref{app:variable-ratio} into the equation derived for $\chi$. It gives
\begin{equation}
 \ddot\Psi_{\rm pub}+2\frac{\dot{\mathcal K}}{\mathcal K}\dot\Psi_{\rm pub}
 +\left(\omega^2-V_{\mathrm{even}}+\frac{\ddot{\mathcal K}}{\mathcal K}\right)\Psi_{\rm pub}=0.
 \label{app:published-variable-equation}
\end{equation}
Thus the literal published variable does not give the same Schr\"odinger normalization under the action and conventions used here. The comparison between the two derivations concerns the canonical operators, not an assumed equality of every intermediate master-variable definition.

\subsection{Symbolic checks and scope of the agreement}
\label{app:symbolic-checks}

Two self-contained symbolic programs implement the reductions in opposite orders. The action program contracts $F_{\mu\nu}F^{\mu\nu}$ and $G^{\mu\rho}F_{\mu\nu}F_\rho{}^\nu$ in coordinates $(t,r,\cos\theta,\phi)$ and performs the angular integral before deriving the radial equations. The field-equation program uses $(t,r,\theta,\phi)$, differentiates the invariant Lagrangian with respect to all six independent field-strength components, and evaluates every component of $\nabla_\mu P^{\mu\nu}$. Each computes the curvature from its metric independently. Neither reads the other program's intermediate expressions or takes a target potential as its input.

The original angular and component checks use $m=0$, $\ell=1,2,3$, with the gauge function retained in the second route. An additional component calculation evaluates all four divergences for $(\ell,m)=(1,1),(2,1),(2,2)$ with both parities present, testing the angular dependence in Eq.~\eqref{app:component-residual-definition}. Exact algebraic checks also test the expanded residual identity, the unfixed radial-action variations, $N_1,N_2$, and the general-background potentials. The general-$\ell,m$ reduction follows from Eq.~\eqref{app:angular-identities}, rather than extrapolation from this finite set. The checks also verify the canonical transformations, the $h$-variable transformation, and the Maxwell and Schwarzschild limits. Finally, the derived RN factors obey
\begin{equation}
 \begin{gathered}
 W_{\mathrm{even}}(\alpha)W_{\mathrm{odd}}(-\alpha)=1,\\
 W_{\mathrm{even}}-W_{\mathrm{odd}}=\frac{x^2}{1-x},
 \end{gathered}
 \label{app:checks}
\end{equation}
where the denominators are nonzero. These identities are consequences and implementation checks of the derived potentials. They do not replace the variational derivations, nor do the regular-variable transformations establish a connection condition across a constitutive zero.
\endgroup

\bibliographystyle{JHEP}
\bibliography{refs}

\end{document}